\documentclass{osa-article}
\usepackage{siunitx}

\def\be{\begin{equation}}
\def\ee{\end{equation}}
\def\bes{\begin{equation*}}
\def\ees{\end{equation*}}

\renewcommand{\vec}[1]{\mathbf{#1}}

\journal{oe}
\articletype{Research Article}
\begin{document}

\title{Brownian dynamics simulations of sphere clusters in optical tweezers}

\author{Wyatt Vigilante, Oscar Lopez, and Jerome Fung\authormark{*}}

\address{Department of Physics and Astronomy, Ithaca College, 953 Danby Road, Ithaca, NY 14850, USA}

\email{\authormark{*}jfung@ithaca.edu} %% email address is required

% \homepage{http:...} %% author's URL, if desired

%%%%%%%%%%%%%%%%%%% abstract %%%%%%%%%%%%%%%%
%% [use \begin{abstract*}...\end{abstract*} if exempt from copyright]

\begin{abstract}
Computationally modeling the behavior of wavelength-sized non-spherical particles in optical tweezers
can give insight into the existence and stability of trapping equilibria as well as
the optical manipulation of such particles more broadly.
Here, we report Brownian dynamics simulations of non-spherical particles that account for detailed optical,
hydrodynamic, and thermal interactions.
We use a $T$-matrix formalism to calculate the optical forces and torques 
exerted by focused laser beams on clusters of wavelength-sized
spheres, and we incorporate detailed diffusion tensors that capture the anisotropic Brownian motion of the clusters.
For two-sphere clusters whose size is comparable to or larger than the wavelength, we observe photokinetic 
effects in elliptically-polarized beams.
We also demonstrate that multiple trapping equilibria exist for a highly asymmetric chiral cluster of seven spheres.
Our simulations may lead to practical suggestions for optical trapping and manipulation as well as a deeper 
understanding of the underlying physics.
\end{abstract}

%%%%%%%%%%%%%%%%%%%%%%%%%%  body  %%%%%%%%%%%%%%%%%%%%%%%%%%
\section{Introduction}

Modeling the behavior of wavelength-sized particles in optical tweezers can 
guide experiments in optical manipulation or assembly.
Doing so requires calculating the optical forces and torques exerted on particles in a focused laser beam.
These calculations are particularly challenging for wavelength-sized particles since 
 neither the Rayleigh approximation nor ray optics can be validly used.
Rather, detailed consideration of the electromagnetic fields incident on and scattered by the particle is necessary
 \cite{jones_optical_2015}.
$T$-matrices provide a well-established formalism for computing and describing electromagnetic scattering by non-spherical
 particles \cite{mishchenko_t-matrix_1996, mishchenko_light_1991, mackowski_calculation_1996}.
To date, however, the use of $T$-matrices to calculate forces and torques in optical tweezers has mainly been limited
to rotationally symmetric particles or particles that are smaller than the wavelength
\cite{borghese_radiation_2006, borghese_optical_2007, borghese_rotational_2007, 
nieminen_optical_2007, nieminen_t-matrix_2011, cao_equilibrium_2012, 
qi_comparison_2014}.

Even so, being able to compute optical forces and torques is not sufficient for understanding the behavior of
non-spherical particles in optical tweezers.
Key questions include the positions, orientations, and stability of any trapping equilibria that may exist.
But as Bui \emph{et al.}~point out, finding equilibria solely from calculations of forces and torques at fixed positions
and orientations is difficult for non-spherical particles because of the size of the parameter space \cite{bui_theory_2017}. 
Consider for example a sphere in a circularly polarized beam.
Rotational symmetry requires that the sphere center lie on
the beam axis in equilibrium.
Moreover, the sphere's orientation is irrelevant.
So, to find the equilibrium trapping position, it is only necessary to calculate the axial component of the trapping force 
as a function of the sphere's position along the beam axis.
No such simplifications are possible for particles lacking symmetry; mapping the trapping landscape for an asymmetric
particle may require considering 3 positional and 3 orientational degrees of freedom \cite{bui_theory_2017}.

An alternative strategy for finding equilibria is to perform dynamical simulations that use computations of optical
forces and torques to model the actual motion of a particle in optical tweezers.
If the goal is only to find trapping equilibria, it is not necessary to carefully consider other interactions that a trapped
particle may experience but that vanish in static equilibrium, such as hydrodynamic resistance.
In contrast, investigating the stability of any trapping equilibria or the dynamics of transitions between multiple
equilibria requires more care. 
Optically-trapped particles suspended in a fluid 
experience Brownian motion involving both hydrodynamic interactions and random thermal interactions.
These can both be highly anisotropic.
Detailed consideration of Brownian motion is also needed 
to explore photokinetic effects arising from the angular momentum carried by an elliptically- or circularly-polarized beam
since they can involve an interplay between optical and thermal interactions
\cite{simpson_first-order_2010, ruffner_optical_2012}.

Here, we report the first Brownian dynamics simulations for wavelength-sized non-spherical particles 
in optical tweezers that 
account for optical interactions, hydrodynamic interactions, and thermal fluctuations in detail.
We combine \emph{mstm}, a well-established package for calculating $T$-matrices of clusters of spheres 
\cite{mackowski_calculation_1996, mackowski_multiple_2011}
with \emph{ott}, a package that can calculate the optical forces and torques on a particle in a focused laser beam 
given a $T$-matrix \cite{nieminen_optical_2007, lenton_ilent2ott_2020}.
Furthermore, we use complete diffusion tensors for the clusters in order to realistically predict
their anisotropic behavior in a fluid.
Our work not only extends the size regime for which optical forces and torques on sphere clusters have previously
been reported but also predicts their detailed motion in optical tweezers for the first time.
While dynamical simulations of particles in optical tweezers have previously been reported 
\cite{simpson_first-order_2010, cao_equilibrium_2012, lenton_optical_2018, bui_theory_2017, volpe_simulation_2013,
armstrong_swimming_2020},
our simulations incorporate both thermal fluctuations and anisotropic hydrodynamic resistance,
and we do not linearize the optical forces and torques.
We validate our simulations by considering single spheres and then explore the remarkably rich dynamical 
effects that occur for clusters of two spheres that are each comparable to or larger than the wavelength of 
the trapping laser.
We then use our simulations to find multiple trapping equilibria for a highly asymmetric cluster
of seven spheres.

\section{Simulation methods}

\subsection{Simulating Brownian motion with external interactions}

\begin{figure}[htbp]
\centering
\includegraphics[width=1.75in]{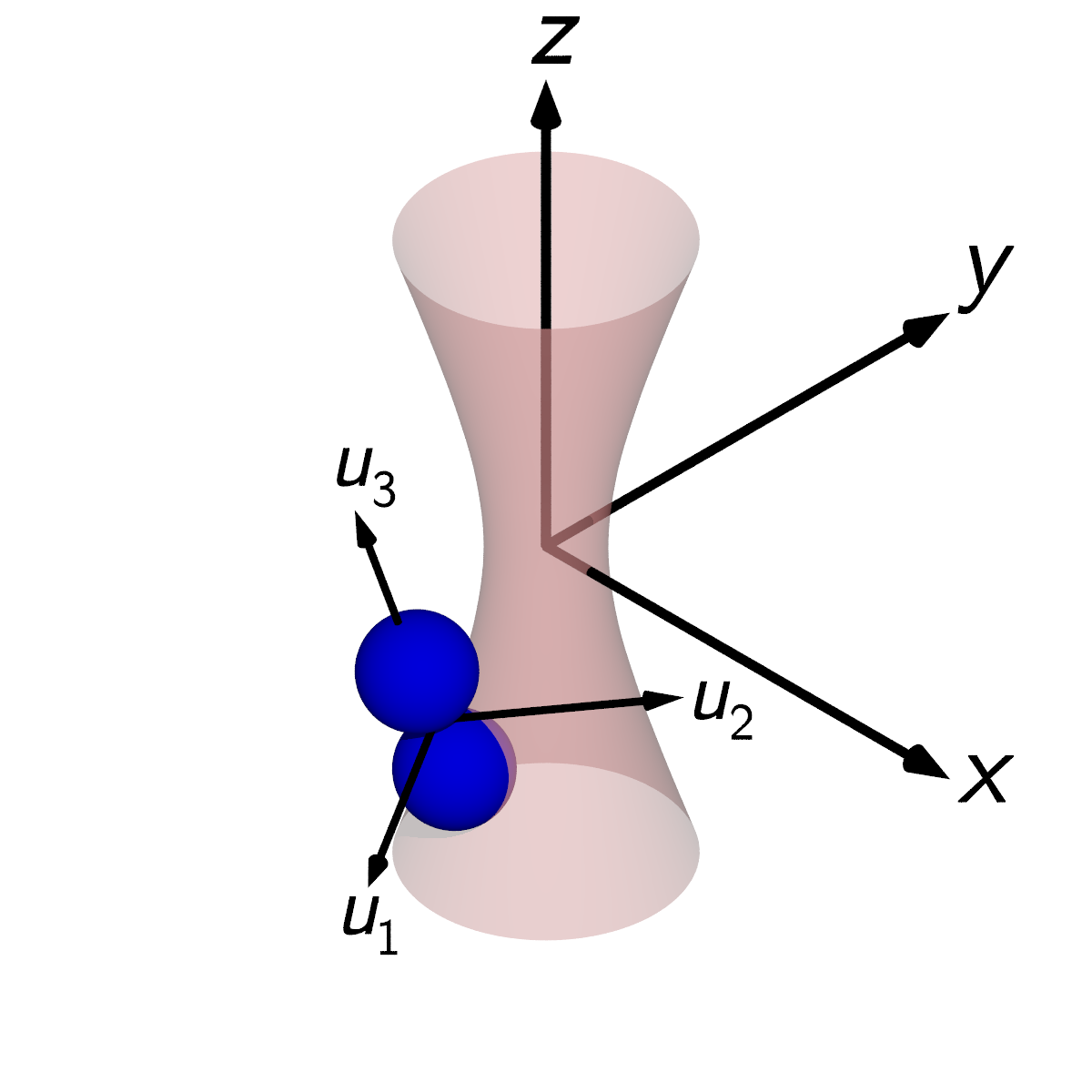}
\caption{\label{fig:cartoon} Coordinate systems used in simulations. In the laboratory coordinate system, a laser beam 
focused at $(0,0,0)$ propagates in the $+z$ direction. In the particle coordinate system, the spheres of a given 
cluster are located at fixed reference positions. The origin of the particle coordinate system is located at the cluster's
center of mass. The orientation of the particle axes in the laboratory coordinate system describes the 
cluster's orientation.}
\end{figure}

% describe particle trajectories
We seek to simulate the trajectory of an arbitrary Brownian particle at temperature $T$ in a fluid of viscosity $\eta$
that is subject to external forces and torques.
We assume that our particles are rigid and move in what we call the ``laboratory frame.''
Consequently, we can describe the position and orientation of our particles by giving the laboratory coordinates of their
center of mass and a rotation matrix.
Specifically, for clusters of spheres, we define a reference configuration in which the position of each sphere
is defined relative to the particle coordinate axes $\vec{u}_1$, $\vec{u}_2$, and $\vec{u}_3$, which we take to be unit vectors.
We choose the origin of the particle coordinate axes to lie at the cluster center of mass (CM).
The components of the particle coordinate axes in the laboratory frame are given by the columns of a $3\times3$ rotation matrix.
Figure \ref{fig:cartoon} illustrates the laboratory and particle coordinate systems.

% describe simulation approach:
% overdamped limit of langevin equation
% diffusion tensor
% generalized force
In the overdamped limit in which the particle's inertia is negligible, we model the trajectory of a Brownian particle
using a finite-difference approach by computing its generalized displacements during time steps of duration $\Delta t$
\cite{fernandes_brownian_2002, volpe_simulation_2013}.
Following Fernandes and Garc\'{i}a de la Torre \cite{fernandes_brownian_2002}, 
we consider the 6-component generalized displacement vector $\Delta \vec{q}$ whose transpose is given by
\be
\Delta \vec{q}^\mathrm{tr} = \left( \Delta x, \Delta y, \Delta z, \Delta \phi_1, \Delta \phi_2, \Delta \phi_3 \right).
\ee
Here, $\Delta x$, $\Delta y$, and $\Delta z$ are displacements of the center of mass
and $\Delta \phi_1$, $\Delta \phi_2$, and $\Delta \phi_3$ are infinitesimal rotations about the particle's axes $\vec{u}_1$, $\vec{u}_2$,
and $\vec{u}_3$.
The generalized displacement $\Delta \vec{q}$ is computed in the particle coordinate system and subsequently needs
to be transformed to laboratory coordinates.
To find $\Delta \vec{q}$ for a given time step, we separately consider two
contributions:
\be \label{eq:Delta_q_decomp}
\Delta \vec{q} = \Delta \vec{q}^{B} + \Delta \vec{q}^\mathrm{ext}. 
\ee
The first term is due to the particle's random Brownian motion, while the second term is due to the external interactions.
The Brownian displacement $\Delta \vec{q}^B$ with components $\Delta q^B_i$ is simulated by generating 
normally distributed random numbers with covariance 
\be \label{eq:Delta_q_b}
\langle \Delta q^B_i \Delta q^B_j \rangle = 2D_{ij} \Delta t,
\ee
where the $D_{ij}$ are the elements of the particle's $6\times6$ diffusion tensor $\mathbf{D}$ \cite{fernandes_brownian_2002}.
The displacement $\Delta \vec{q}^\mathrm{ext}$ due to an external force $\vec{F}_\mathrm{ext}$ and torque $\vec{n}_\mathrm{ext}$
is given by
\be \label{eq:Delta_q_ext}
\Delta \vec{q}^\mathrm{ext} = \left( \frac{\Delta t}{k_BT} \right) \mathbf{D}\cdot\mathcal{F}, 
\ee
where $\vec{F}_\mathrm{ext}$ and $\vec{n}_\mathrm{ext}$ have been combined into the 6-component vector $\mathcal{F}$
\cite{fernandes_brownian_2002}.
Our main computational challenges 
are computing the external forces and torques $\mathcal{F}$ and finding the diffusion tensor $\mathbf{D}$.

\subsection{Calculating optical forces and torques with $T$-matrices}
While the approach in Eqs.~(\ref{eq:Delta_q_b}) and (\ref{eq:Delta_q_ext}) is relevant to any external interaction, we now focus
on the optical forces and torques exerted by optical tweezers.
We consider a laser beam of vacuum wavelength $\lambda_0$ in a medium of refractive index $n_\mathrm{med}$ that is focused 
by an objective lens with a given numerical aperture (NA).

Calculating the optical forces and torques is fundamentally a light scattering problem.
As we briefly summarize here, we expand the position-dependent
electric field of the incident beam $\vec{E}_\mathrm{inc}(k\vec{r})$ in vector spherical wave functions (VSWFs) 
\cite{nieminen_optical_2014}:
\be
\vec{E}_\mathrm{inc}(k\vec{r}) =  \sum_{n=1}^\infty \sum_{m=-n}^{n} a_{nm}\mathrm{Rg}\vec{M}_{nm}(k\vec{r}) + b_{nm} \mathrm{Rg}\vec{N}_{nm}(k\vec{r}).
\ee
Here, $\mathrm{Rg}\vec{M}_{nm}(k\vec{r})$ and $\mathrm{Rg}\vec{N}_{nm}(k\vec{r})$ are solutions of the vector
Helmholtz equation that are finite at the origin and $k$ is the beam's wavenumber.
(See Nieminen \emph{et al.} and references therein for details \cite{nieminen_optical_2014}.)
We use \emph{ott} to truncate the expansion and calculate the expansion coefficients $a_{nm}$ and $b_{nm}$ 
for a focused Gaussian beam using
an overdetermined least squares fit \cite{nieminen_multipole_2003, nieminen_optical_2007}.
The fitting procedure is necessary because a Gaussian beam is not a solution to the vector Helmholtz equation.

We similarly expand the scattered field $\vec{E}_\mathrm{sca}(k\vec{r})$ in VSWFs \cite{nieminen_optical_2014}:
\be
\vec{E}_\mathrm{sca}(k\vec{r}) =  \sum_{n=1}^\infty \sum_{m=-n}^{n} p_{nm}\vec{M}^{(1)}_{nm}(k\vec{r}) + q_{nm} \vec{N}^{(1)}_{nm}(k\vec{r}).
\ee
The $\vec{M}^{(1)}_{nm}$ and $\vec{N}^{(1)}_{nm}$ VSWFs asymptotically behave as outgoing spherical waves.
The details of the scattering process are described by the $T$-matrix $\mathbf{T}$, which
relates the scattered coefficients to the incident coefficients:
\be \label{eq:t_matrix_mult}
\begin{pmatrix}
p_{nm} \\ q_{nm}
\end{pmatrix}
=
\mathbf{T}
\begin{pmatrix}
a_{nm} \\ b_{nm}
\end{pmatrix},
\ee
where we combine the incident and scattered field coefficients into single column vectors.
\emph{ott} can natively calculate $T$-matrices for homogenous spheres, whose elements are the Lorenz-Mie coefficients.
However, \emph{ott} cannot calculate $T$-matrices for sphere clusters.
Instead, we use the multisphere superposition code \emph{mstm} developed by Mackowski and Mishchenko 
\cite{mackowski_multiple_2011}.
Our code calls \emph{mstm}, reads the $T$-matrix it generates, and then uses \emph{ott} to perform the matrix multiplication in 
Eq.~(\ref{eq:t_matrix_mult}) to find the scattered field expansion coefficients $p_{mn}$ and $q_{mn}$.
Subsequently, \emph{ott} determines the optical forces and torques from the 
field expansion coefficients 
$a_{nm}$, $b_{nm}$, $p_{nm}$, and $q_{nm}$ \cite{nieminen_optical_2007}. 
In this approach, the computationally challenging light scattering problem only needs to be solved once in a general way when
computing the $T$-matrix; subsequent force and torque calculations are fast.

\subsection{Finding diffusion tensors}
Brownian dynamics simulations require knowing the diffusion tensor $\mathbf{D}$ of 
a particle.
Finding $\mathbf{D}$ requires solving the Stokes equations for creeping flow around the particle.
For spheres of radius $a$, $\mathbf{D}$ is diagonal with two unique, nonzero elements:
the translational diffusion constant  $D_t = k_BT / (6\pi\eta a)$ and 
the rotational diffusion constant $D_r = k_BT / (8\pi\eta a^3)$.
For clusters of two identical spheres, or dimers, $\mathbf{D}$ is also diagonal and has four unique elements.
$D_{t,\parallel}$ describes translational diffusion along the dimer's long axis while $D_{t,\perp}$ describes translational
diffusion perpendicular to the long axis. Similarly, $D_{r,\parallel}$ describes rotational diffusion about the long axis 
and $D_{r,\perp}$ describes rotational diffusion about the other two axes.
We use the analytic solution of Nir \& Acrivos to find $\mathbf{D}$ for dimers \cite{nir_creeping_1973}. 
However, for more complex sphere clusters, numerical methods are needed.
We use the Fortran program BEST, which implements a boundary element solution to the Stokes equations 
over an arbitrary triangulated surface \cite{aragon_precise_2004}.
As recommended, we perform repeated BEST calculations using increasing numbers of triangles and 
extrapolate to the limit of an infinite number of triangles.
By comparing the results of BEST for single spheres and dimers to the analytic results, we estimate that using
BEST results in uncertainties in the diffusion tensor elements of no more than 1\%.

\subsection{Implementation}
We implement our Brownian dynamics algorithms in the package \emph{brownian\_ot}, which is freely available on Github
\cite{jerome_jeromefungbrownian_ot_nodate}.
Our code integrates both a Matlab package (\emph{ott}) and a Fortran 90 package (\emph{mstm}) in a single interface.
To make this interfacing straightforward, we implement \emph{brownian\_ot} in Python.

The initial steps in simulating a given particle in a beam of a specified wavelength are determining the particle's diffusion
tensor, determining the particle's $T$-matrix for scattering of light of that wavelength, and determining the beam's
VSWF expansion coefficients. These calculations only need to be done once.
Thereafter, we simulate the trajectory of a particle beginning from an arbitrary initial position and orientation as follows:
\begin{enumerate}
 \item Use \emph{ott} to compute the optical force and torque $\mathcal{F}$ on the particle in the laboratory coordinate system.
 \item Correct the optical torque to be relative to the particle's center of diffusion \cite{harvey_coordinate_1980}
 and transform all elements of $\mathcal{F}$ to the particle coordinate system.
 \item Use Eq.~(\ref{eq:Delta_q_ext}) to calculate $\Delta \vec{q}^\mathrm{ext}$ in the particle coordinate system.
 \item Use Eq.~(\ref{eq:Delta_q_b}) to calculate $\Delta \vec{q}^B$ in the particle coordinate system.
 \item Compute $\Delta \vec{q} = \Delta \vec{q}^\mathrm{ext} + \Delta \vec{q}^B$ [Eq.~(\ref{eq:Delta_q_decomp})].
 \item Transform the CM displacement (the spatial components of $\Delta \vec{q}$) to the laboratory coordinate system and update
 the CM position.
 \item Calculate an unbiased rotation corresponding to the angular components of $\Delta \vec{q}$
\cite{beard_unbiased_2003}
 and update the particle's orientation by rotation composition.
 \item Repeat.
\end{enumerate}
Internally, our orientation calculations use unit quaternions due to their compact storage and numerical stability.

% Describe typical runtime configuration
While our code runs on personal computers under Macintosh, Windows, or Linux operating systems, 
we typically run longer simulations using the Comet cluster
at the San Diego Supercomputing Center, which is part of the Extreme Science and Engineering Discovery Environment (XSEDE)
\cite{towns_xsede_2014}.
The primary reason to use Comet is that we can run tens of independent simulations simultaneously, 
either to build up statistics or to explore parameter space.
Run times depend primarily on the number of time steps to be simulated but also on the particle size
since larger particles generally require a larger maximum order $n_\mathrm{max}$ in the truncation of the VSWF expansions for 
the incident and scattered fields.
On Comet, simulating $3\times10^6$ steps for a 1-\si{\micro\meter}-diameter sphere takes approximately \SI{8e4}{\second},
and simulating a trajectory of the same length for a cluster of seven $0.8$-\si{\micro\meter}-diameter spheres takes approximately 
\SI{1e5}{\second}.

\subsection{Simulation parameters}
Throughout what follows, we consider a 1064-nm-wavelength beam focused by a 1.2 NA objective lens.
Except when noted, the beam has a power of \SI{5}{\milli\watt} and is left-circularly-polarized (LCP) with Jones vector
$(1, i)$.
We also assume that the beam propagates in an isotropic medium with $n_\mathrm{med}=1.33$, which corresponds to water.
We assume that the medium has viscosity $\eta = \SI{1e-3}{\pascal\second}$
and that the particles are at $T=\SI{295}{\kelvin}$.
We consider non-absorbing particles with two different refractive indices: silica (Si; $n=1.45$) and polystyrene (PS; 
$n=1.58$).
However, our code can be easily applied to particles with other refractive indices, including absorbing particles 
such as metallic nanospheres.

We choose a simulation time step $\Delta t = \SI{1e-5}{\second}$ throughout our simulations. 
Because optical forces and torques are strongly dependent on particle position and orientation, we choose the time step 
to be small enough that none of our particles translates a significant fraction of the wavelength or rotates through a significant angle
during a single step.
At the same time, $\Delta t$ is large enough that ballistic motion is negligible \cite{volpe_simulation_2013, huang_direct_2011}. 
Finally, except when noted, all simulations begin with the particle centers of mass at the origin and the particle reference
axes aligned with the laboratory frame axes.

\section{Results and discussion}

\subsection{Spheres}

\begin{figure}[htbp]
\centering
\includegraphics{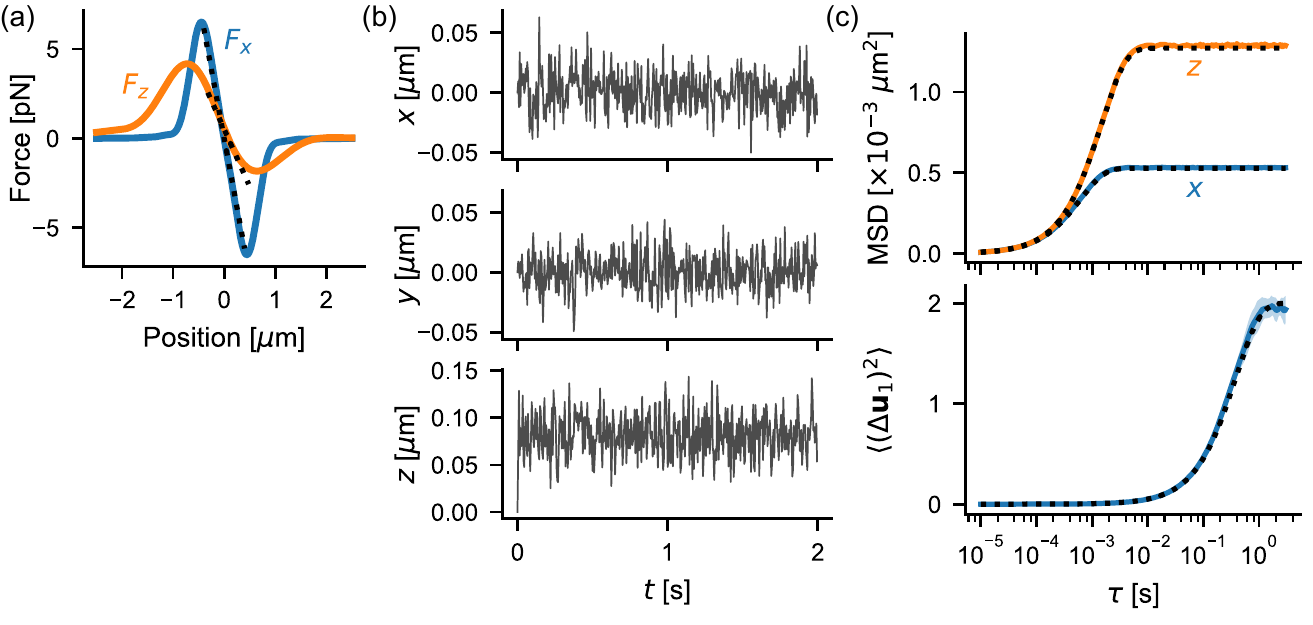}
\caption{\label{fig:sphere} (a) Optical forces on a 1-\si{\micro\meter}-diameter PS sphere in water in a 5 mW LCP beam as a 
function of particle position. $F_z$ is calculated as a function of $z$ at $x=y=0$. $F_x$ is calculated as a function of $x$
at $y=0$ and $z=z_\mathrm{eq}$ (at which $F_z = 0$). Dashed lines indicate linear approximation to optical forces near equilibrium
giving stiffnesses $\kappa_x$ and $\kappa_z$.
(b) First \SI{2}{\second} of a 30-s-long simulated trajectory. 
(c) Laboratory-frame MSD and $\vec{u}_1$ MSAD averaged from 5 independent trajectories with
identical initial conditions. 
Shaded regions indicate standard deviations. Dotted lines are the predictions of Eqs.~(\ref{eq:MSD}) and (\ref{eq:MSAD}).}
\end{figure}

To validate our simulations, we consider the behavior of a 1-\si{\micro\meter}-diameter PS sphere.
The optical forces as a function of position
are approximately linear near equilibrium (Fig.~\ref{fig:sphere}(a)):
\begin{gather} \label{eq:linear_restoring_force}
F_x = -\kappa_x x, \nonumber \\
F_z = -\kappa_z (z - z_\mathrm{eq}),
\end{gather}
where $\kappa_x$ and $\kappa_z$ are the stiffnesses in the $x$ and $z$ directions, respectively.
Because of radiation pressure, $F_z = 0$ at $z=z_\mathrm{eq}$ with $z_\mathrm{eq}>0$ rather than at $z =0$.
The first \SI{2}{\second} of a simulated sphere trajectory shows that the sphere exhibits Brownian fluctuations about equilibrium
(Fig.~\ref{fig:sphere}(b)).

Mean-squared displacements (MSDs) calculated from the simulated trajectories demonstrate that the simulations perform
as expected. The MSD for the coordinate $x_i$ is defined as $\mathrm{MSD}=\langle \left[ x_i(t+\tau) - x_i(t) \right]^2\rangle$.
We plot the $x$ and $z$ MSDs calculated from our simulations in Fig.~\ref{fig:sphere}(c).
Since we know the stiffnesses $\kappa_x$ and $\kappa_z$ from the force calculations in Fig.~\ref{fig:sphere}(a) and also know
the sphere radius, we can also predict the MSDs \emph{a priori}.
Solving the Langevin equation shows that a sphere of radius $a$ moving in one dimension subject to a linear restoring force with 
stiffness $\kappa$ has the MSD \cite{jones_optical_2015}
\be \label{eq:MSD}
\mathrm{MSD} = \frac{2k_BT}{\kappa}\left[  1 - \exp\left( -\frac{\kappa \tau}{\gamma}\right)  \right],
\ee
where the friction coefficient is given by $\gamma = 6\pi\eta a$.
The dotted lines in the upper panel of Fig.~\ref{fig:sphere}(c) show that the predictions of Eq.~(\ref{eq:MSD})
agree well with the values calculated from the trajectories.

In addition, because there are no significant optical torques, the sphere should undergo free rotational
diffusion. We can similarly characterize the rotational dynamics of the sphere by calculating the mean-squared angular
displacement (MSAD) of any of the reference axes in the laboratory frame. The MSAD for axis $\vec{u}_i$
is defined as $\langle (\Delta\vec{u}_i)^2\rangle = \langle \left|\vec{u}_i(t+\tau) - \vec{u}_i(t)\right|^2 \rangle$. 
The lower panel of Fig.~\ref{fig:sphere}(c) shows the MSAD for $\vec{u}_1$ since, for a sphere, all three axes are equivalent.
A sphere with rotational diffusion constant $D_r$ has an MSAD given by 
\cite{doi_theory_1988, fung_holographic_2013}
\be \label{eq:MSAD}
\langle (\Delta \vec{u}_i )^2\rangle = 2\left[ 1 - \exp(-2D_r\tau  )  \right].
\ee
The agreement between the prediction of Eq.~(\ref{eq:MSAD}) (the dotted line in the lower panel of Fig.~\ref{fig:sphere}(c))
and the values determined from the simulated trajectories shows that the sphere indeed undergoes free rotational diffusion as
expected.

\subsection{Dimers}

\begin{figure}[htbp]
\centering
\includegraphics{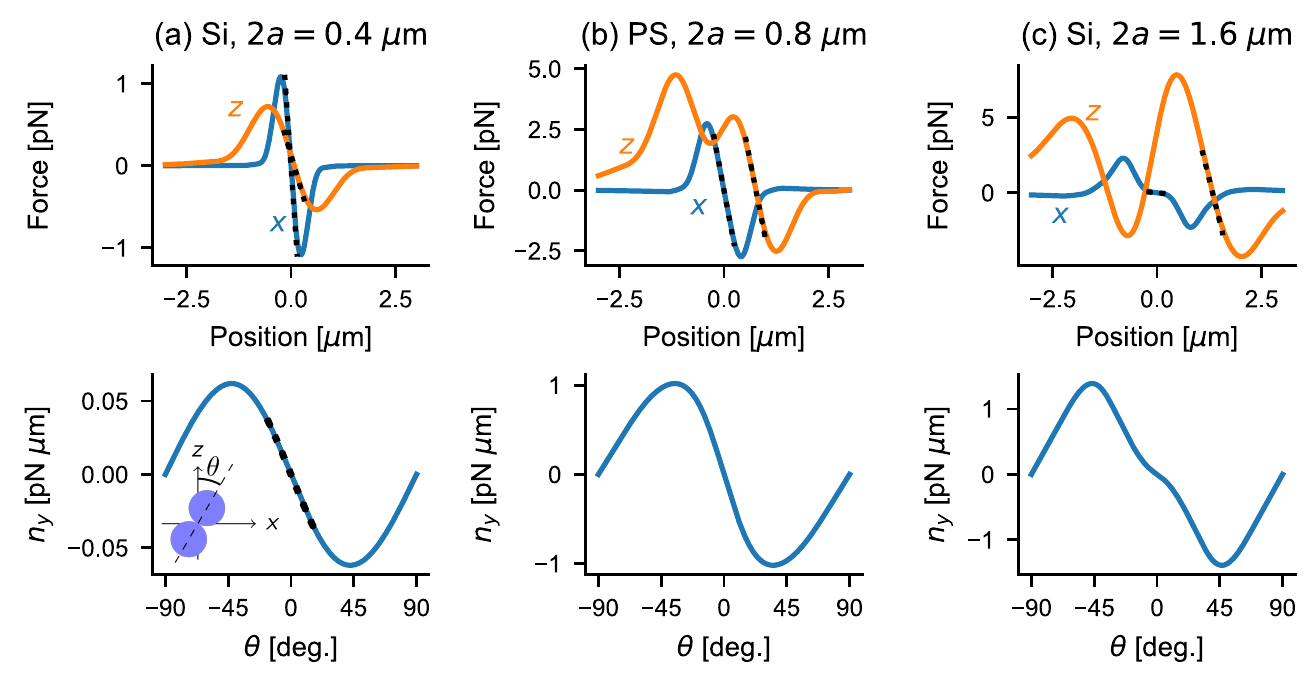}
\caption{\label{fig:dimer_forces} Optical forces and torques on small, intermediate, and large dimers in a 5 mW LCP beam.
All force calculations are shown with the dimer long axis aligned with the $z$ axis.
$F_x$ and $n_y$ are calculated at $z=z_\mathrm{eq}$.
In (c), the equilibrium position with $z_{eq}>0$ is chosen.
Inset in (a) shows definition of $\theta$. 
Dotted lines in force graphs indicate linear approximations near equilibrium. 
The dotted line in the torque graph for (a) shows a small-angle fit to Eq.~(\ref{eq:restoring_torque}).}
\end{figure}

We next consider the simplest multi-sphere cluster: a dimer consisting of two identical spheres.
Figure \ref{fig:cartoon} shows the reference orientation for a dimer; we choose the long axis to be the
$\vec{u}_3$ axis.

In order to verify that the dimers we consider can indeed be optically trapped, we compute the optical forces and torques on 
several characteristic dimers (Fig.~\ref{fig:dimer_forces}).
We perform all force calculations with the dimer long axis aligned with the direction of beam propagation.
We observe three regimes.
First, for a dimer whose constituent spheres are smaller than the wavelength of light, which we term ``small,'' both 
$F_x$ and $F_z$ resemble the corresponding calculations for spheres (Fig.~\ref{fig:dimer_forces}(a)).
In particular, there is a linear regime near equilibrium and an equilibrium height $z_\mathrm{eq}>0$.
In addition, the lateral stiffness $\kappa_x$ is larger than the axial stiffness $\kappa_z$.
These results are similar to those reported by Borghese \emph{et al}.~for 220-nm-diameter PS spheres \cite{borghese_optical_2007}.
Second, we consider an ``intermediate'' dimer whose spheres are comparable in size to the wavelength of light 
in the surrounding medium.
For the dimer of 0.8-\si{\micro\meter}-diameter spheres shown in Fig.~\ref{fig:dimer_forces}(b),
we observe a qualitatively different $F_z$ graph with two maxima.
Also, in contrast to the small dimer, the equilibrium trap stiffnesses $\kappa_x$ and $\kappa_z$ are comparable.
Third, we consider a ``large'' dimer consisting of spheres that are larger than the wavelength,
such as the 1.6-\si{\micro\meter}-diameter silica spheres in Fig.~\ref{fig:dimer_forces}(c).
Here, there are two heights $z$ where $F_z=0$. In addition,  
the trap is stiffer axially than laterally: $\kappa_z>\kappa_x$.

The optical torques experienced by small, intermediate, and large dimers also differ.
Since the incident beam is axisymmetric, we consider without loss of generality
rotating the dimer by $\theta$ about the $y$ axis and compute the $y$ component of the optical torque $n_y$.
In all three cases, there is a restoring torque resulting in a stable angular equilibrium at $\theta=0$.
Notably, for the small dimer in Fig.~\ref{fig:dimer_forces}(a), the torque is approximately proportional to $\sin(2\theta)$:
\be \label{eq:restoring_torque}
n_y = -\kappa_r\sin(2\theta),
\ee
where $\kappa_r$ is a rotational stiffness. 
However, Eq.~(\ref{eq:restoring_torque}) poorly describes the torques on intermediate and large dimers.

\begin{figure}[htbp]
\centering
\includegraphics{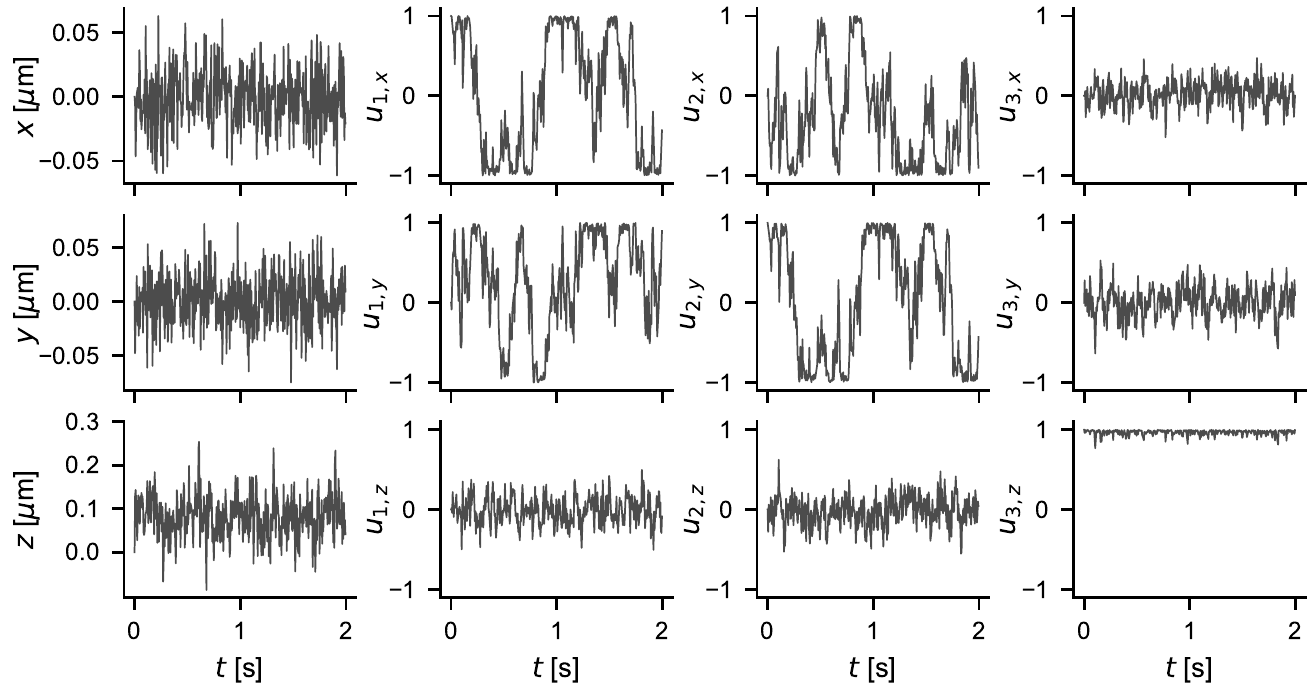}
\caption{\label{fig:small_traj} First \SI{2}{s} of the CM trajectory and rotation matrix elements for a small 
dimer composed of 0.4-\si{\micro\meter}-diameter silica spheres in a 5 mW LCP trap. 
The complete 30-s-long trajectory is shown in \textcolor{urlblue}{Visualization 1}.}
\end{figure}

\begin{figure}[htbp]
\centering
\includegraphics[width=0.8\textwidth]{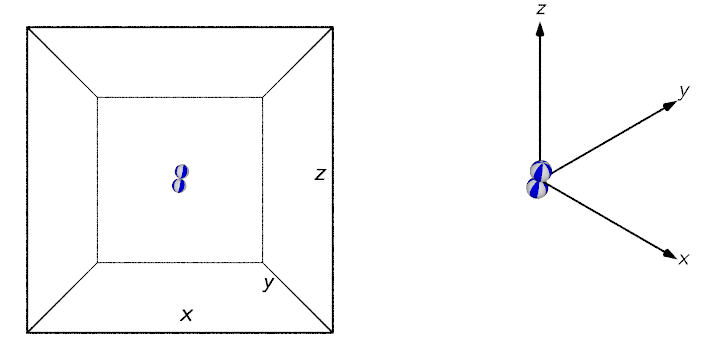}
\caption{\label{fig:movie} Rendering of an optically-trapped dimer composed of 0.4-\si{\micro\meter}-diameter silica spheres from 
\textcolor{urlblue}{Visualization 1}, which shows the trajectory of Fig.~\ref{fig:small_traj}.
The stripes are drawn to enable visualizing rotations about all axes.
Left: front view. The laser propagates upwards and is focused at the center of the box, 
whose sides are \SI{6}{\micro\meter} long.
Right: perspective view. The axes are \SI{3.5}{\micro\meter} long. 
All subsequent visualizations have the same scale, and all visualizations are slowed $5\times$ from real time.}
\end{figure}

Small, intermediate, and large dimers exhibit qualitatively different dynamical behavior in optical tweezers.
Small dimers exhibit positional fluctuations similar to those for a trapped sphere and 
orientational fluctuations that would be expected for a particle experiencing a restoring torque.
The beginning of a simulation trajectory for a small dimer is shown in Fig.~\ref{fig:small_traj}, where
we plot the CM coordinates and the laboratory-frame coordinates of all 3 particle-frame axes.
\textcolor{urlblue}{Visualization 1} shows a rendering of the complete trajectory, and Fig.~\ref{fig:movie} is a still frame from the rendering.
Once again, the CM position fluctuates about $x=y=0$ and $z=z_\mathrm{eq}$.
Unlike spheres, however, the restoring torque keeps the dimer axis $\vec{u}_3$ pointing near the laboratory $+z$ axis:
$u_{3x}$ and $u_{3y}$ fluctuate near 0 and $u_{3z}$ fluctuates near 1.
The horizontal components of $\vec{u}_{1}$ and $\vec{u}_2$ range fully between -1 and +1, indicating that the dimer is free to
rotate about its long axis. 

\begin{figure}[htbp]
\centering
\includegraphics{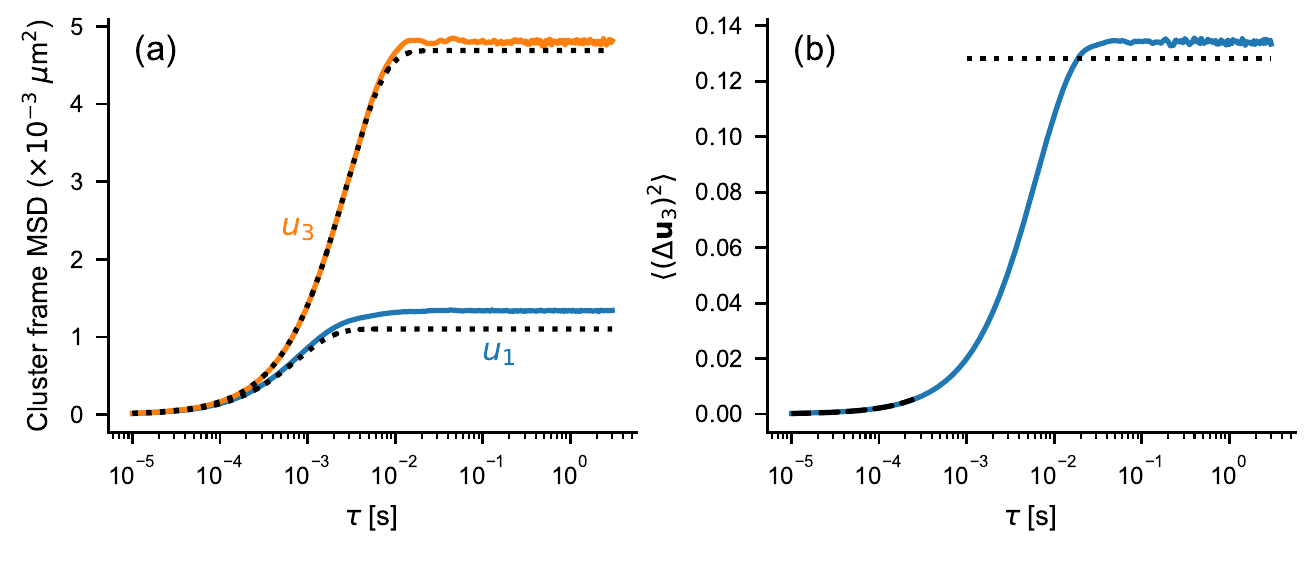}
\caption{\label{fig:dimer_msds}
Cluster-frame MSDs and $\vec{u}_3$ MSAD for dimer of 0.4-\si{\micro\meter}-diameter silica spheres averaged from 5 
independent trajectories. 
Predictions of Eq.~(\ref{eq:MSD}): dotted lines in (a). Prediction of Eq.~(\ref{eq:msad_limit}) for MSAD 
limiting value: dotted line in (b).
Prediction of Eq.~(\ref{eq:MSAD}) for short-time behavior of MSAD: dashed line in (b).}
\end{figure}

We analyze the behavior shown in the trajectory by computing MSDs and MSADs (Fig.~\ref{fig:dimer_msds}).
Because the diffusion tensor is anisotropic, we compute MSDs in the particle frame by resolving laboratory-frame displacements
into components along the particle axes \cite{han_brownian_2006, fung_holographic_2013}. 
We determine trap stiffnesses $\kappa_x$ and $\kappa_z$ from the linear behavior of $F_x$ and $F_z$ near equilibrium shown in 
Fig.~\ref{fig:dimer_forces}(a).
% TODO: think about how to present this 
We also extract translational drag coefficients $\gamma_\parallel = k_BT/D_{t,\parallel}$ and $\gamma_\perp = k_BT/D_{t,\perp}$ 
from the diffusion tensor.
We can therefore compare the predictions of Eq.~(\ref{eq:MSD}) to the MSDs calculated from the trajectories.

We also calculate the MSAD for the long axis $\vec{u}_3$ since 
the orientation of this axis can be observed in experiments using techniques such as holographic microscopy 
\cite{fung_measuring_2011}. 
At short lag times $\tau$, the optical torques do not affect the dimer: the MSAD is diffusive and is well-described by 
Eq.~(\ref{eq:MSAD}) with $D_r = D_{r,\perp}$ (Fig.~\ref{fig:dimer_msds}(b)).
At longer times, however, the effects of the optical tweezers become significant.
Even approximating the torque to be proportional to $\sin(2\theta)$ [Eq.~(\ref{eq:restoring_torque})], we cannot analytically 
solve the Einstein-Smoluchowski equation in order to determine the $\vec{u}_3$ MSAD for arbitrary lag times. 
But assuming that the particle is in thermal equilibrium, we can compute the limiting value of the MSAD as $\tau\rightarrow\infty$.
(See Supplemental Document for details.)
The limiting value is given by
\be \label{eq:msad_limit}
\lim_{\tau \rightarrow \infty} \langle( \Delta \vec{u}_3)^2\rangle = 2\left[ 1 - \frac{1}{4\beta\kappa_r} 
\left( \frac{\exp(\beta\kappa_r) - 1}{\exp(\beta \kappa_r)F(\sqrt{\beta\kappa_r})}\right)^2    \right],
\ee
where $\beta = 1/(k_BT)$ and $F(x)$ is Dawson's integral.
Since we can determine $\kappa_r$ by fitting the torque calculations shown in Fig.~\ref{fig:dimer_forces}(a),
we can plot the predictions of Eq.~(\ref{eq:msad_limit}) in Fig.~\ref{fig:dimer_msds}(b).

The results in Fig.~\ref{fig:dimer_msds} all differ from the analytical predictions by several percent.
The discrepancies suggest that the translational and orientational degrees of freedom cannot be considered independently
of each other.
In particular, the optical forces on the dimer depend not only on position but also on orientation.
Similarly, the optical torques depend not only on orientation but also on position.
Moreover, the optical torque is only approximately described by Eq.~(\ref{eq:restoring_torque}).

\begin{figure}[htbp]
\centering
\includegraphics{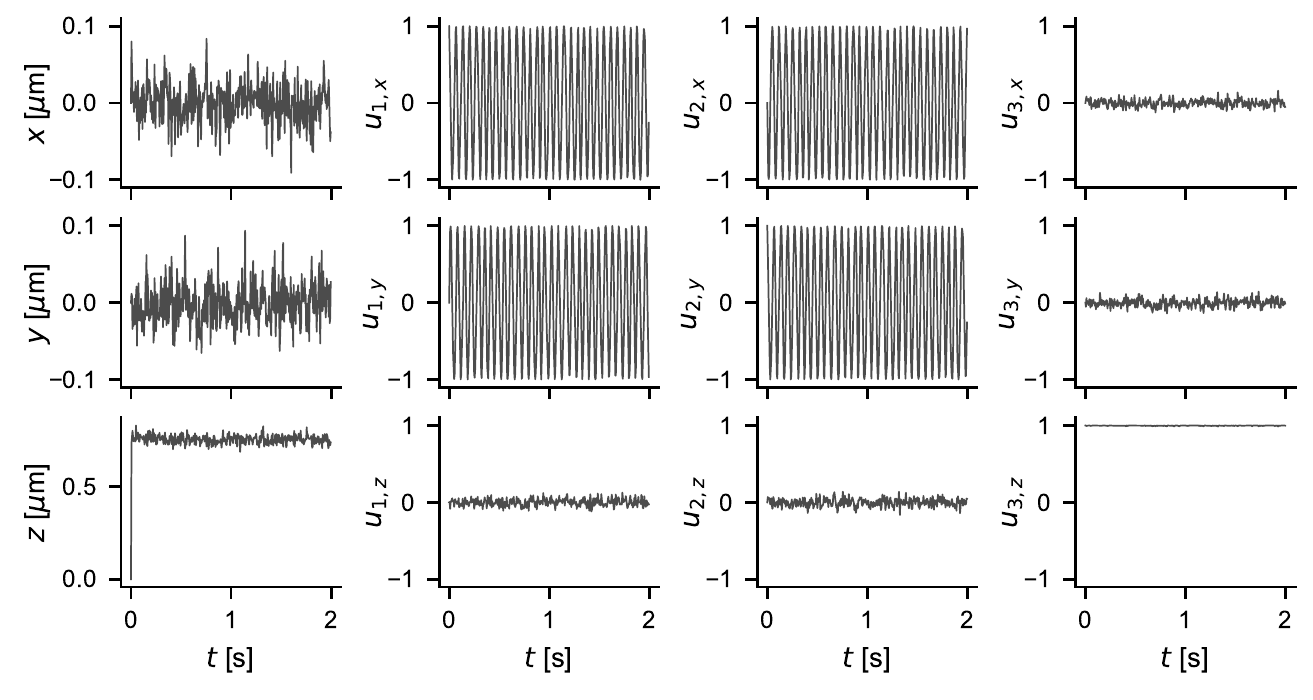}
\caption{\label{fig:medium_traj} First \SI{2}{s} of the CM trajectory and rotation matrix elements for an intermediate dimer 
composed of 0.8-\si{\micro\meter}-diameter PS spheres in a 5 mW LCP trap. The complete 30-s-long trajectory is shown in 
\textcolor{urlblue}{Visualization 2}. The dimer exhibits deterministic rotation about its long axis.}
\end{figure}

\begin{figure}[htbp]
\centering
\includegraphics{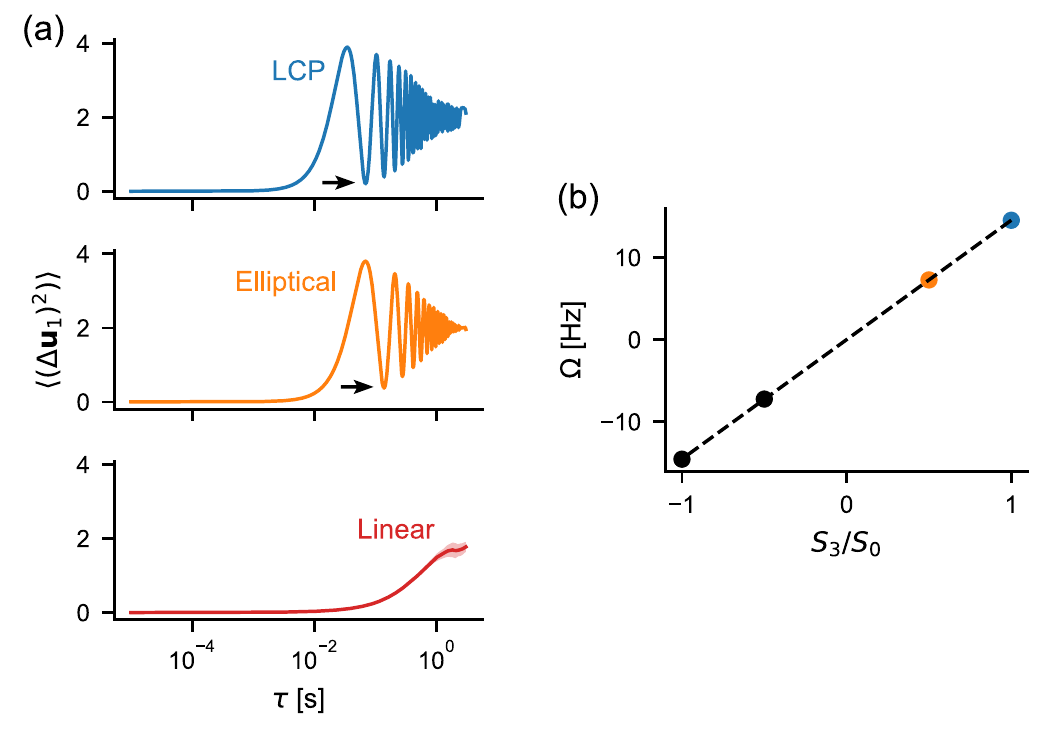}
\caption{ \label{fig:dimer_rotation}
(a) $\vec{u}_1$ MSADs for dimer of 0.8-\si{\micro\meter}-diameter PS spheres in 5 mW beams with left circular polarization, elliptical polarization with Jones vector $(1/\sqrt{2}, \exp(i\pi/6)/\sqrt{2})$, and linear polarization with Jones vector 
$(1/\sqrt{2}, 1/\sqrt{2})$. MSADs are averaged from 5 trajectories.
The period of the dimer's deterministic rotation is determined from the first minimum of the MSADs (arrows).
No rotation is observed for the linearly polarized beam.
(b) Rotation frequency $\Omega$ as a function of the ratio of Stokes vector elements $S_3/S_0$. 
The blue and orange points correspond to the MSADs shown in (a). 
The sign of $\Omega$ corresponds to the direction of rotation. Dashed line: linear fit. }
\end{figure}

The trajectory of the intermediate dimer in Fig.~\ref{fig:medium_traj} shows a marked difference from that of the small dimer:
the dimer rotates deterministically about its long axis.
This can be clearly seen in the behavior of $\vec{u}_1$ and $\vec{u}_2$ and in \textcolor{urlblue}{Visualization 2}.
The rotation is polarization-dependent: the dimer does not rotate when the incident beam is linearly polarized.
Moreover, the direction of rotation reverses when the beam is right-circularly polarized (\textcolor{urlblue}{Visualization 3}).
We attribute this behavior to photokinetic spin-curl effects \cite{ruffner_optical_2012, yevick_photokinetic_2017}.
While their analysis is only strictly valid in the Rayleigh limit, Ruffner and Grier predicted theoretically 
and demonstrated experimentally that the rotation frequency for particles experiencing spin-curl torques is proportional to 
$S_3/S_0$, where $(S_0, S_1, S_2, S_3)$ is the Stokes vector describing the incident polarization \cite{ruffner_optical_2012}. 
We thus perform additional simulations with elliptically-polarized beams and 
determine the periods of rotation, averaged over 5 trajectories, by computing the $\vec{u}_1$ MSAD and
locating the first minimum (Fig.~\ref{fig:dimer_rotation}(a)).
The rotation frequencies $\Omega$ we observe are indeed proportional to $S_3/S_0$ (Fig.~\ref{fig:dimer_rotation}(b)).
Here, a positive $\Omega$ corresponds to a counterclockwise rotation and a negative $\Omega$ corresponds to a counterclockwise
rotation.

\begin{figure}[htbp]
\centering
\includegraphics{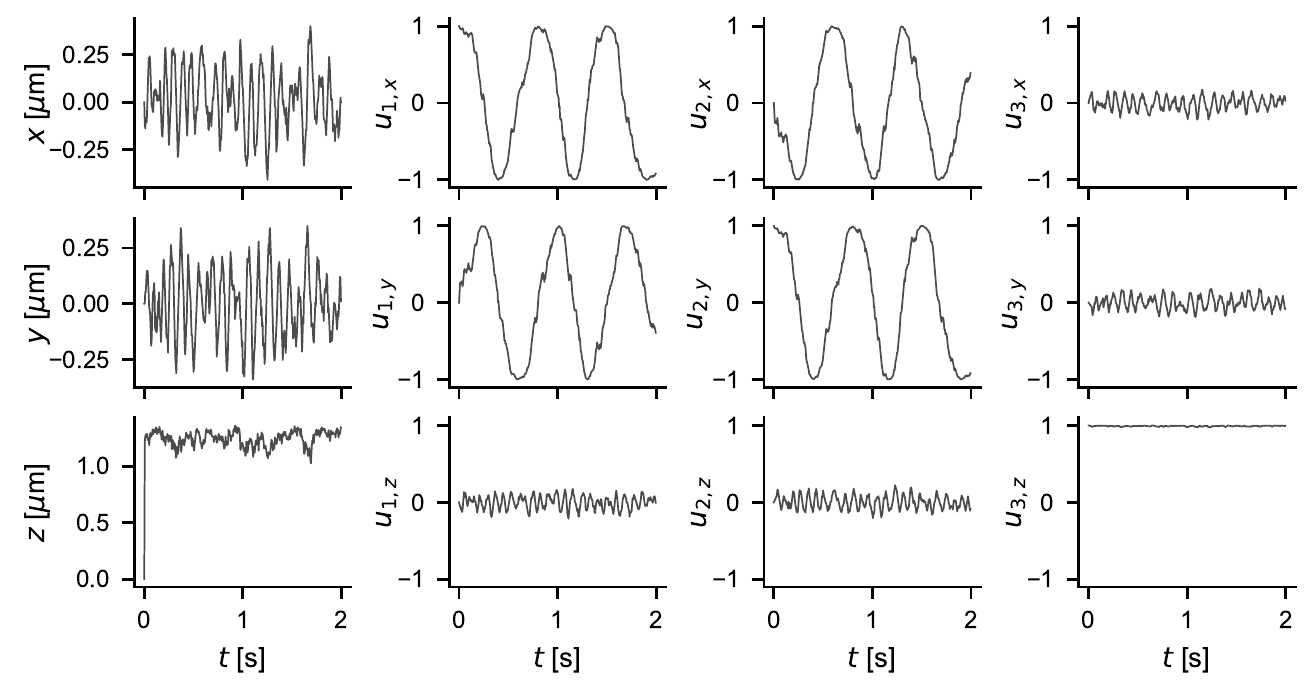}
\caption{\label{fig:large_traj} First \SI{2}{s} of the CM trajectory and rotation matrix elements for a large dimer 
composed of 1.6-\si{\micro\meter}-diameter silica spheres in a 5 mW LCP trap. The complete 30-s-long trajectory is shown in 
\textcolor{urlblue}{Visualization 4}. The dimer exhibits both a slower deterministic rotation about its long axis ($\vec{u}_3$) 
as well as a faster wobble of its long axis.}
\end{figure}

For the large dimer, we observe a similar deterministic rotation about the dimer axis
(Fig.~\ref{fig:large_traj} and \textcolor{urlblue}{Visualization 4}).
However, there is an additional effect: the particle undergoes a wobble that is faster than the axial rotation.
The effects of the wobble can be seen in the $z$ component of $\vec{u}_1$ and $\vec{u}_2$, 
in the lateral components of $\vec{u}_3$, and in the lateral position of the center of mass.
Both the axial rotation and the wobble reverse direction when the beam is 
right-circularly polarized (\textcolor{urlblue}{Visualization 5}).
While experimentally detecting the axial rotation described in Fig.~\ref{fig:dimer_rotation} for an optically homogeneous
medium dimer would be challenging, it would be straightforward to detect the wobble we predict here.
We plan to explore these effects both computationally and experimentally in the future.

\subsection{Chiral 7-sphere cluster}

\begin{figure}[htbp]
\centering
\includegraphics{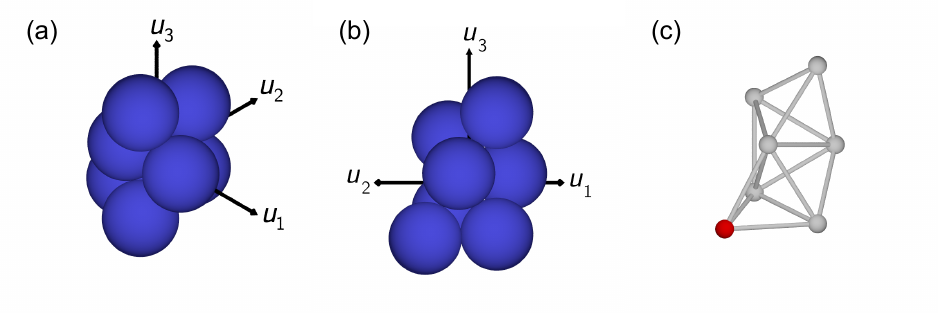}
\caption{\label{fig:chiral7_ref} Reference orientation for chiral 7-sphere cluster. 
(a) and (b): Space-filling models viewed from two perspectives. 
(c): Ball-and-stick model viewed from same perspective as (b). The red sphere causes the cluster to be chiral.}
\end{figure}

Finally, our Brownian dynamics simulations reveal the existence of multiple trapping equilibria for a highly asymmetric
sphere cluster.
The 7-sphere cluster shown in Fig.~\ref{fig:chiral7_ref} has no axes of rotational symmetry
or planes of mirror symmetry, but it is the smallest rigid sphere packing that exhibits chirality \cite{arkus_deriving_2011}.
We simulate clusters of 0.8-\si{\micro\meter}-diameter silica spheres held in a beam that is 
linearly polarized in the $x$ direction.
As in the previous simulations, the cluster begins with its center of mass at the origin and in its reference orientation.

\begin{figure}[htbp]
\centering
\includegraphics{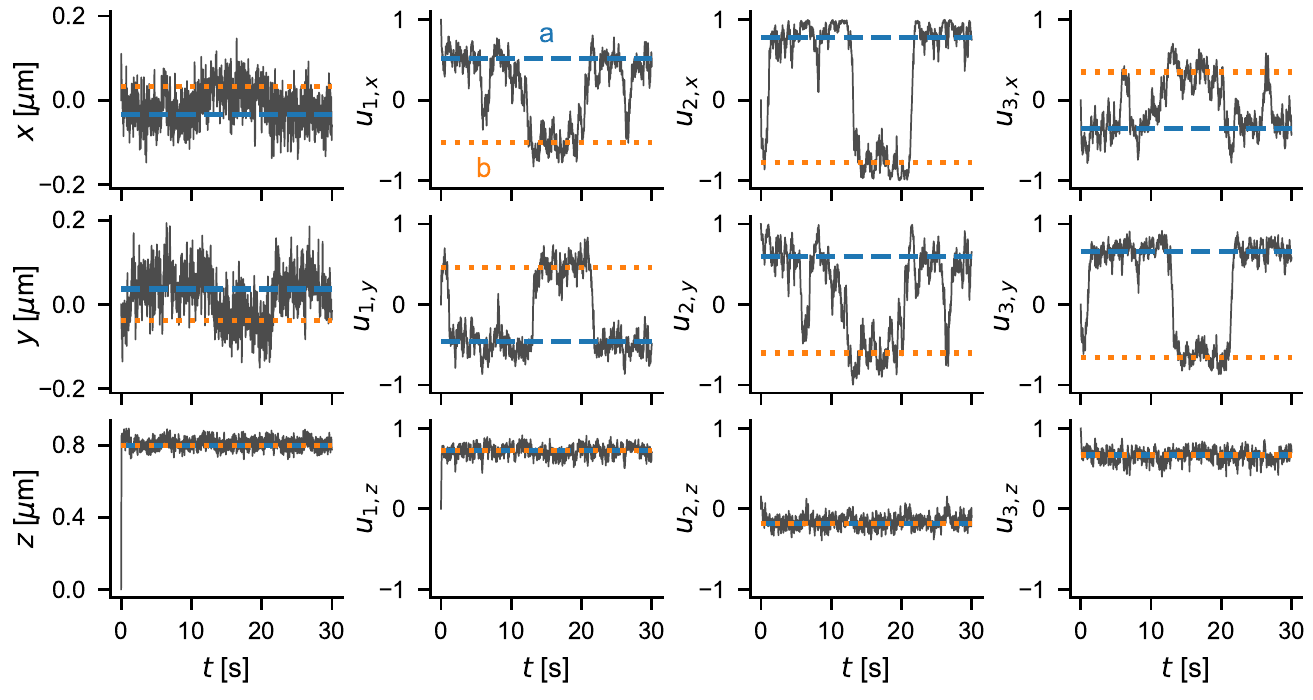}
\caption{\label{fig:chiral7_ab} CM and rotation matrix for a chiral 7-sphere cluster of
0.8-\si{\micro\meter}-diameter silica spheres trapped in a \SI{5}{\milli\watt} horizontally-polarized beam.
(See \textcolor{urlblue}{Visualization 6}.)
Dashed lines indicate equilibria shown in Fig.~\ref{fig:chiral7_equilibria}(a)-(b).}
\end{figure}

Rather than fluctuating or exhibiting deterministic motion about a \emph{single} equilibrium, the simulated trajectory 
fluctuates about two different orientations (Fig.~\ref{fig:chiral7_ab} and \textcolor{urlblue}{Visualization 6}). 
This is most clearly visible in the behavior of the $x$ and $y$ components of the cluster-frame axes, 
although the two orientations also have different values of the CM position.

\begin{figure}[htbp]
\centering
\includegraphics{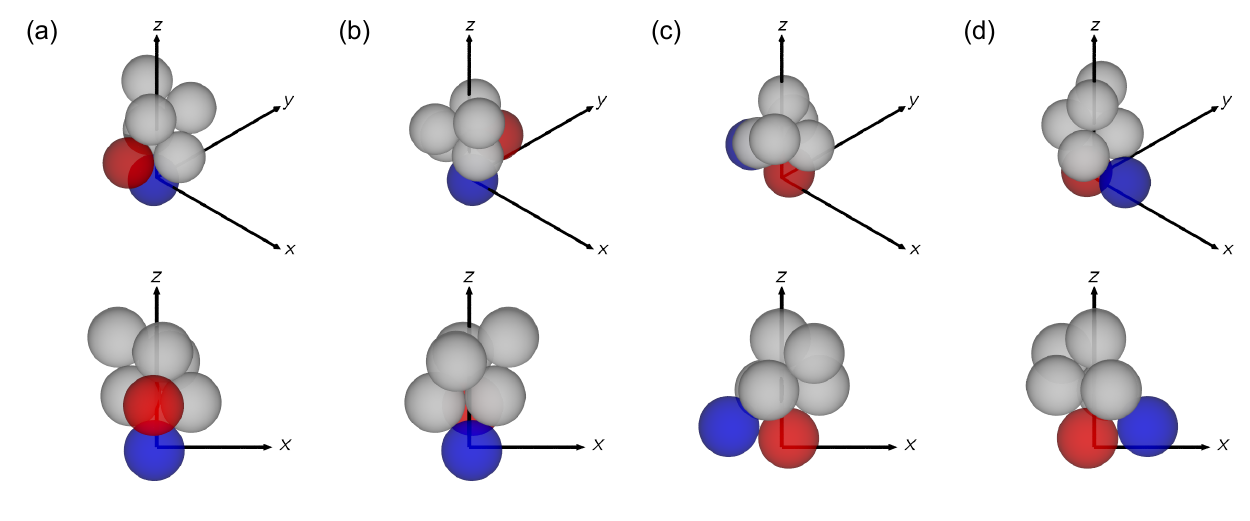}
\caption{\label{fig:chiral7_equilibria}
Equilibrium orientations indicated by dashed lines in Figs.~\ref{fig:chiral7_ab} and \ref{fig:chiral7_cd}. In orientations (a) and (b), 
which are rotated by approximately $180^\circ$ about the $z$ axis, the trap focus is on the sphere 
highlighted in blue. In orientations (c) and (d), also rotated by $180^\circ$ about the $z$ axis, the trap focus is on the sphere
highlighted in red.}
\end{figure}

In order to determine whether the observed plateaus in the trajectory (e.g., between $t=14$ and \SI{19}{\second}) 
in fact correspond to equilibria, we estimate candidate equilibrium positions and orientations from the plateaus
and perform additional athermal simulations starting from the candidate equilibria.
To estimate the equilibrium position, we average each CM coordinate in a plateau.
To estimate the equilibrium orientation, we use the quaternion averaging algorithm
of Markley and co-workers \cite{markley_averaging_2007}.
We then perform additional simulations with the cluster starting from the candidate equilibria but without thermal fluctuations,
and we confirm that the resulting trajectories approach asymptotic values.
The refined equilibria obtained from the limiting values
are shown in the horizontal lines of Fig.~\ref{fig:chiral7_ab} and in Fig.~\ref{fig:chiral7_equilibria}(a)-(b).
The two equilibria here correspond to a rotation of approximately $180^\circ$ about the $z$ axis.

\begin{figure}[htbp]
\centering
\includegraphics{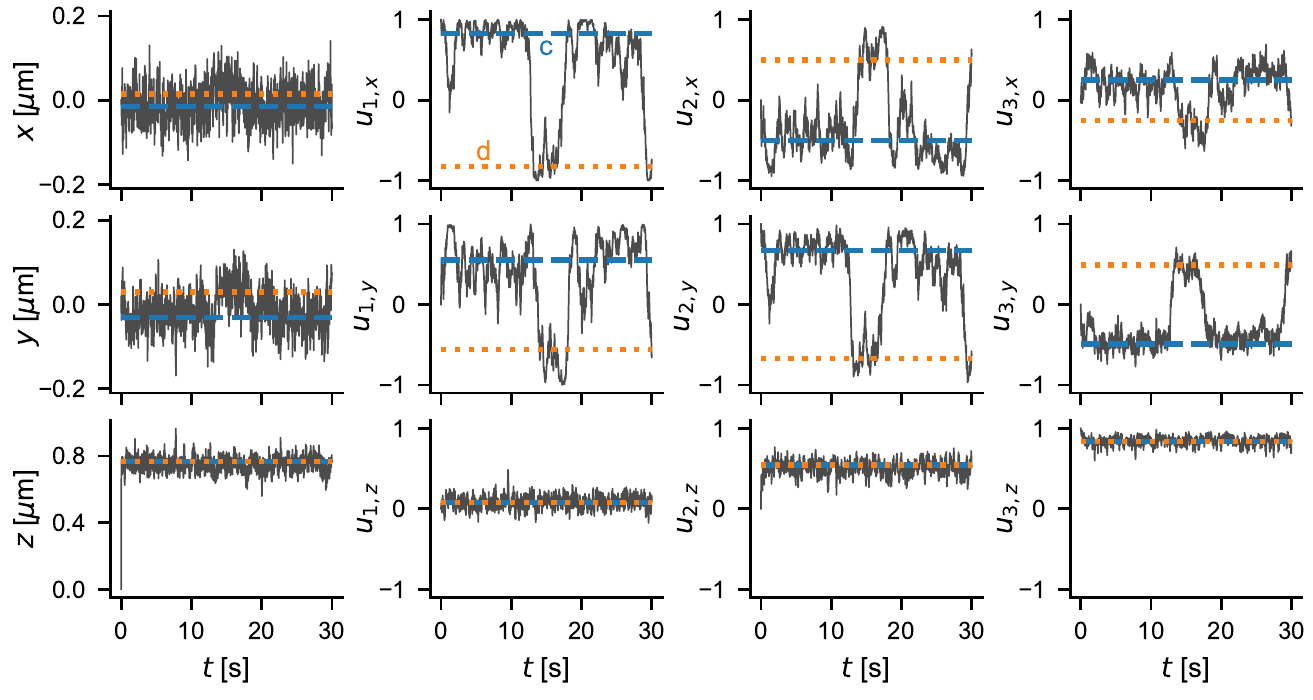}
\caption{\label{fig:chiral7_cd} As in Fig.~\ref{fig:chiral7_ab}, but for a different simulation. 
 (See \textcolor{urlblue}{Visualization 7}.) 
 Dashed lines indicate equilibria shown in Fig.~\ref{fig:chiral7_equilibria}(c)-(d).} 
\end{figure}

Remarkably, in a different simulation with the \emph{same} initial conditions, we observe a different set of candidate equilibria
(Fig.~\ref{fig:chiral7_cd} and \textcolor{urlblue}{Visualization 7}).
We again perform additional simulations without thermal fluctuations and show the refined equilibria in Figs.~\ref{fig:chiral7_cd}
and \ref{fig:chiral7_equilibria}(c)-(d).
Once again, the two equilibria in this trajectory correspond to a $180^\circ$ rotation about the $z$ axis. 
However, the equilibria of Fig.~\ref{fig:chiral7_equilibria}(c)-(d) involve a different sphere lying at the beam focus than in 
Fig.~\ref{fig:chiral7_equilibria}(a)-(b).

These results demonstrate the importance of considering thermal fluctuations in simulations of particles in optical tweezers.
Here, thermal fluctuations allow the cluster to explore its trapping landscape and reveal multiple equilibria that we did not
expect to find \emph{a priori}. 
There may be additional trapping equilibria besides the ones we have reported.
We intend to explore the trapping of this cluster in more detail as well as the transitions between
equilibria.

\section{Conclusions}
Combining $T$-matrix-based calculations of optical interactions with a model of anisotropic Brownian motion
has resulted in dynamical simulations that realistically capture the behavior of complex, wavelength-sized sphere 
clusters in optical tweezers.
We have observed rich photokinetic effects even for the simplest case, clusters of two identical isotropic spheres, when
the individual spheres are comparable to or larger than the wavelength.
Our simulations have also demonstrated that multiple trapping equilibria exist for a wavelength-sized 7-sphere cluster 
with no symmetry.
Finding this cluster's trapping equilibria solely from calculations of optical interactions at fixed positions and orientations
would have been computationally challenging.
Moreover, our incorporation of thermal fluctuations allowed us to more easily explore the trapping landscape.
The fully general, on-demand computations of optical interactions needed for these simulations were only possible because of the
speed of the $T$-matrix-based calculations.

Our work suggests two intriguing directions for future inquiry.
First, systematically investigating effects such as the photokinetic axial rotation and wobble
we have observed for intermediate and large dimers would be worthwhile.
Our simulations allow continuous variation of parameters such as the particle size and refractive
index and could usefully complement experimental investigations in which the particle properties 
cannot be as easily tuned.
Moreover, since we can in principle determine the electromagnetic fields everywhere 
from the VSWF expansion of the incident beam and the $T$-matrix, more detailed investigation into 
the physical mechanism of these effects might be possible.
This may be worthwhile since theoretical analyses based on the 
Rayleigh approximation cannot be strictly valid for particles of these sizes \cite{yevick_photokinetic_2017}.

Second, it is also possible to extend this work to other systems.
Using \emph{ott}, we could investigate particle dynamics in beams that are not Gaussian, 
including Bessel beams \cite{arlt_optical_2001}
and Laguerre-Gaussian beams \cite{simpson_mechanical_1997, 
simpson_optical_2009}, both of which can carry orbital angular momentum.
In addition, the generality of the $T$-matrix approach and the modularity of our code
could allow us to consider other particles.
\emph{ott} natively supports $T$-matrix computations for small spheroids, and
other packages reliably calculate the $T$-matrix for spheroids \cite{somerville_smarties_2016}
or other axisymmetric particles \cite{mishchenko_capabilities_1998}.
The discrete dipole approximation, which is implemented in \emph{ott},
 could also enable $T$-matrices to be computed for arbitrarily-shaped
particles \cite{mackowski_discrete_2002, loke_discrete-dipole_2011}.
The broad generality as well as detail of the Brownian dynamics simulations we have introduced here 
may make them useful both for guiding or interpreting experiments on optically-trapped 
particles as well as for gaining further insight into the underlying physics of optical trapping and manipulation.

\section*{Funding}
This work used the Extreme Science and Engineering Discovery Environment (XSEDE), which is supported by National 
Science Foundation grant number ACI-1548562.
This work used the XSEDE resource Comet at the San Diego Supercomputing Center through allocation PHY190049.
O.~L. was supported by the Summer Scholars program of the Ithaca College School of Humanities \& Sciences.

\section*{Acknowledgments}
We thank Isaac Lenton, Miranda Holmes-Cerfon, and Sergio Aragon for helpful discussions.

\section*{Disclosures}

The authors declare no conflicts of interest.

\vspace{1em}
\noindent
See Supplement 1 for supporting content.

%%%%%%%%%%%%%%%%%%%%%%% References %%%%%%%%%%%%%%%%%%%%%%%%%

%%%%%%%%%% If using BibTeX:
\bibliography{bd}

\begin{thebibliography}{10}
\newcommand{\enquote}[1]{``#1''}

\bibitem{jones_optical_2015}
P.~H. Jones, O.~M. Maragò, and G.~Volpe, \emph{Optical {Tweezers}:
  {Principles} and {Applications}} (Cambridge University Press, Cambridge,
  2015).

\bibitem{mishchenko_t-matrix_1996}
M.~I. Mishchenko, L.~D. Travis, and D.~W. Mackowski, \enquote{T-matrix
  computations of light scattering by nonspherical particles: {A} review,}
  {\protect\JournalTitle{Journal of Quantitative Spectroscopy and Radiative
  Transfer}} \textbf{55}, 535--575 (1996).

\bibitem{mishchenko_light_1991}
M.~I. Mishchenko, \enquote{Light scattering by randomly oriented axially
  symmetric particles,} {\protect\JournalTitle{JOSA A}} \textbf{8}, 871--882
  (1991).

\bibitem{mackowski_calculation_1996}
D.~W. Mackowski and M.~I. Mishchenko, \enquote{Calculation of the {T} matrix
  and the scattering matrix for ensembles of spheres,}
  {\protect\JournalTitle{JOSA A}} \textbf{13}, 2266--2278 (1996).

\bibitem{borghese_radiation_2006}
F.~Borghese, P.~Denti, R.~Saija, and M.~A. Iatì, \enquote{Radiation torque on
  nonspherical particles in the transition matrix formalism,}
  {\protect\JournalTitle{Optics Express}} \textbf{14}, 9508--9521 (2006).

\bibitem{borghese_optical_2007}
F.~Borghese, P.~Denti, R.~Saija, and M.~A. Iatì, \enquote{Optical trapping of
  nonspherical particles in the {T}-matrix formalism,}
  {\protect\JournalTitle{Optics Express}} \textbf{15}, 11984--11998 (2007).

\bibitem{borghese_rotational_2007}
F.~Borghese, P.~Denti, R.~Saija, and M.~A. Iatì, \enquote{On the rotational
  stability of nonspherical particles driven by the radiation torque,}
  {\protect\JournalTitle{Optics Express}} \textbf{15}, 8960--8971 (2007).

\bibitem{nieminen_optical_2007}
T.~A. Nieminen, V.~L.~Y. Loke, A.~B. Stilgoe, G.~Knöner, A.~M. Brańczyk,
  N.~R. Heckenberg, and H.~Rubinsztein-Dunlop, \enquote{Optical tweezers
  computational toolbox,} {\protect\JournalTitle{Journal of Optics A: Pure and
  Applied Optics}} \textbf{9}, S196--S203 (2007).

\bibitem{nieminen_t-matrix_2011}
T.~A. Nieminen, V.~L.~Y. Loke, A.~B. Stilgoe, N.~R. Heckenberg, and
  H.~Rubinsztein-Dunlop, \enquote{T-matrix method for modelling optical
  tweezers,} {\protect\JournalTitle{Journal of Modern Optics}} \textbf{58},
  528--544 (2011).

\bibitem{cao_equilibrium_2012}
Y.~Cao, A.~B. Stilgoe, L.~Chen, T.~A. Nieminen, and H.~Rubinsztein-Dunlop,
  \enquote{Equilibrium orientations and positions of non-spherical particles in
  optical traps,} {\protect\JournalTitle{Optics Express}} \textbf{20},
  12987--12996 (2012).

\bibitem{qi_comparison_2014}
X.~Qi, T.~A. Nieminen, A.~B. Stilgoe, V.~L.~Y. Loke, and H.~Rubinsztein-Dunlop,
  \enquote{Comparison of {T}-matrix calculation methods for scattering by
  cylinders in optical tweezers,} {\protect\JournalTitle{Optics Letters}}
  \textbf{39}, 4827--4830 (2014).

\bibitem{bui_theory_2017}
A.~A.~M. Bui, A.~B. Stilgoe, I.~C.~D. Lenton, L.~J. Gibson, A.~V. Kashchuk,
  S.~Zhang, H.~Rubinsztein-Dunlop, and T.~A. Nieminen, \enquote{Theory and
  practice of simulation of optical tweezers,} {\protect\JournalTitle{Journal
  of Quantitative Spectroscopy and Radiative Transfer}} \textbf{195}, 66--75
  (2017).

\bibitem{simpson_first-order_2010}
S.~H. Simpson and S.~Hanna, \enquote{First-order nonconservative motion of
  optically trapped nonspherical particles,} {\protect\JournalTitle{Physical
  Review E}} \textbf{82}, 031141 (2010).

\bibitem{ruffner_optical_2012}
D.~B. Ruffner and D.~G. Grier, \enquote{Optical {Forces} and {Torques} in
  {Nonuniform} {Beams} of {Light},} {\protect\JournalTitle{Physical Review
  Letters}} \textbf{108}, 173602 (2012).

\bibitem{mackowski_multiple_2011}
D.~W. Mackowski and M.~I. Mishchenko, \enquote{A multiple sphere {T}-matrix
  {Fortran} code for use on parallel computer clusters,}
  {\protect\JournalTitle{Journal of Quantitative Spectroscopy and Radiative
  Transfer}} \textbf{112}, 2182--2192 (2011).

\bibitem{lenton_ilent2ott_2020}
I.~C.~D. Lenton, \enquote{ott: Optical tweezers toolbox,} {github} (2018)
  [retrieved 31 August 2020], \url{https://github.com/ilent2/ott}.

\bibitem{lenton_optical_2018}
I.~C.~D. Lenton, A.~A.~M. Bui, T.~A. Nieminen, A.~B. Stilgoe, and
  H.~Rubinsztein-Dunlop, \enquote{Optical tweezers toolbox: full dynamics
  simulations for particles of all sizes,} {\protect\JournalTitle{Proc. SPIE}}
  \textbf{10723}, 107232B (2018).

\bibitem{volpe_simulation_2013}
G.~Volpe and G.~Volpe, \enquote{Simulation of a {Brownian} particle in an
  optical trap,} {\protect\JournalTitle{American Journal of Physics}}
  \textbf{81}, 224--230 (2013).

\bibitem{armstrong_swimming_2020}
D.~J. Armstrong, T.~A. Nieminen, A.~B. Stilgoe, A.~V. Kashchuk, I.~C.~D.
  Lenton, and H.~Rubinsztein-Dunlop, \enquote{Swimming force and behavior of
  optically trapped micro-organisms,} {\protect\JournalTitle{Optica}}
  \textbf{7}, 989--994 (2020).

\bibitem{fernandes_brownian_2002}
M.~X. Fernandes and J.~García de~la Torre, \enquote{Brownian {Dynamics}
  {Simulation} of {Rigid} {Particles} of {Arbitrary} {Shape} in {External}
  {Fields},} {\protect\JournalTitle{Biophysical Journal}} \textbf{83},
  3039--3048 (2002).

\bibitem{nieminen_optical_2014}
T.~A. Nieminen, N.~du~Preez-Wilkinson, A.~B. Stilgoe, V.~L.~Y. Loke, A.~A.~M.
  Bui, and H.~Rubinsztein-Dunlop, \enquote{Optical tweezers: {Theory} and
  modelling,} {\protect\JournalTitle{Journal of Quantitative Spectroscopy and
  Radiative Transfer}} \textbf{146}, 59--80 (2014).

\bibitem{nieminen_multipole_2003}
T.~A. Nieminen, H.~Rubinsztein-Dunlop, and N.~R. Heckenberg, \enquote{Multipole
  expansion of strongly focussed laser beams,} {\protect\JournalTitle{Journal
  of Quantitative Spectroscopy and Radiative Transfer}} \textbf{79-80},
  1005--1017 (2003).

\bibitem{nir_creeping_1973}
A.~Nir and A.~Acrivos, \enquote{On the creeping motion of two arbitrary-sized
  touching spheres in a linear shear field,} {\protect\JournalTitle{Journal of
  Fluid Mechanics}} \textbf{59}, 209--223 (1973).

\bibitem{aragon_precise_2004}
S.~Aragon, \enquote{A precise boundary element method for macromolecular
  transport properties,} {\protect\JournalTitle{Journal of Computational
  Chemistry}} \textbf{25}, 1191--1205 (2004).

\bibitem{jerome_jeromefungbrownian_ot_nodate}
J.~Fung, \enquote{brownian\_ot,} {github} (2020) [retrieved 31 August 2020],
  \url{https://github.com/jeromefung/brownian_ot}.

\bibitem{harvey_coordinate_1980}
S.~Harvey and J.~García de~la Torre, \enquote{Coordinate {Systems} for
  {Modeling} the {Hydrodynamic} {Resistance} and {Diffusion} {Coefficients} of
  {Irregularly} {Shaped} {Rigid} {Macromolecules},}
  {\protect\JournalTitle{Macromolecules}} \textbf{13}, 960--964 (1980).

\bibitem{beard_unbiased_2003}
D.~A. Beard and T.~Schlick, \enquote{Unbiased {Rotational} {Moves} for
  {Rigid}-{Body} {Dynamics},} {\protect\JournalTitle{Biophysical Journal}}
  \textbf{85}, 2973--2976 (2003).

\bibitem{towns_xsede_2014}
J.~Towns, T.~Cockerill, M.~Dahan, I.~Foster, K.~Gaither, A.~Grimshaw,
  V.~Hazlewood, S.~Lathrop, D.~Lifka, G.~D. Peterson, R.~Roskies, J.~R. Scott,
  and N.~Wilkins-Diehr, \enquote{{XSEDE}: {Accelerating} {Scientific}
  {Discovery},} {\protect\JournalTitle{Computing in Science Engineering}}
  \textbf{16}, 62--74 (2014).

\bibitem{huang_direct_2011}
R.~Huang, I.~Chavez, K.~M. Taute, B.~Lukić, S.~Jeney, M.~G. Raizen, and E.-L.
  Florin, \enquote{Direct observation of the full transition from ballistic to
  diffusive {Brownian} motion in a liquid,} {\protect\JournalTitle{Nature
  Physics}} \textbf{7}, 576--580 (2011).

\bibitem{doi_theory_1988}
M.~Doi and S.~F. Edwards, \emph{The {Theory} of {Polymer} {Dynamics}}
  (Clarendon Press, 1988).

\bibitem{fung_holographic_2013}
J.~Fung and V.~N. Manoharan, \enquote{Holographic measurements of anisotropic
  three-dimensional diffusion of colloidal clusters,}
  {\protect\JournalTitle{Physical Review E}} \textbf{88}, 020302 (2013).

\bibitem{han_brownian_2006}
Y.~Han, A.~M. Alsayed, M.~Nobili, J.~Zhang, T.~C. Lubensky, and A.~G. Yodh,
  \enquote{Brownian {Motion} of an {Ellipsoid},}
  {\protect\JournalTitle{Science}} \textbf{314}, 626--630 (2006).

\bibitem{fung_measuring_2011}
J.~Fung, K.~E. Martin, R.~W. Perry, D.~M. Kaz, R.~McGorty, and V.~N. Manoharan,
  \enquote{Measuring translational, rotational, and vibrational dynamics in
  colloids with digital holographic microscopy,} {\protect\JournalTitle{Optics
  Express}} \textbf{19}, 8051--8065 (2011).

\bibitem{yevick_photokinetic_2017}
A.~Yevick, D.~J. Evans, and D.~G. Grier, \enquote{Photokinetic analysis of the
  forces and torques exerted by optical tweezers carrying angular momentum,}
  {\protect\JournalTitle{Philosophical Transactions of the Royal Society A:
  Mathematical, Physical and Engineering Sciences}} \textbf{375}, 20150432
  (2017).

\bibitem{arkus_deriving_2011}
N.~Arkus, V.~N. Manoharan, and M.~P. Brenner, \enquote{Deriving {Finite}
  {Sphere} {Packings},} {\protect\JournalTitle{SIAM Journal on Discrete
  Mathematics}} \textbf{25}, 42 (2011).

\bibitem{markley_averaging_2007}
F.~L. Markley, Y.~Cheng, J.~L. Crassidis, and Y.~Oshman, \enquote{Averaging
  {Quaternions},} {\protect\JournalTitle{Journal of Guidance, Control, and
  Dynamics}} \textbf{30}, 1193--1197 (2007).

\bibitem{arlt_optical_2001}
J.~Arlt, V.~Garces-Chavez, W.~Sibbett, and K.~Dholakia, \enquote{Optical
  micromanipulation using a {Bessel} light beam,} {\protect\JournalTitle{Optics
  Communications}} \textbf{197}, 239--245 (2001).

\bibitem{simpson_mechanical_1997}
N.~B. Simpson, K.~Dholakia, L.~Allen, and M.~J. Padgett, \enquote{Mechanical
  equivalence of spin and orbital angular momentum of light: an optical
  spanner,} {\protect\JournalTitle{Optics Letters}} \textbf{22}, 52--54 (1997).

\bibitem{simpson_optical_2009}
S.~H. Simpson and S.~Hanna, \enquote{Optical angular momentum transfer by
  {Laguerre}-{Gaussian} beams,} {\protect\JournalTitle{JOSA A}} \textbf{26},
  625--638 (2009).

\bibitem{somerville_smarties_2016}
W.~R.~C. Somerville, B.~Auguié, and E.~C. Le~Ru, \enquote{smarties:
  {User}-friendly codes for fast and accurate calculations of light scattering
  by spheroids,} {\protect\JournalTitle{Journal of Quantitative Spectroscopy
  and Radiative Transfer}} \textbf{174}, 39--55 (2016).

\bibitem{mishchenko_capabilities_1998}
M.~I. Mishchenko and L.~D. Travis, \enquote{Capabilities and limitations of a
  current {FORTRAN} implementation of the {T}-matrix method for randomly
  oriented, rotationally symmetric scatterers,} {\protect\JournalTitle{Journal
  of Quantitative Spectroscopy and Radiative Transfer}} \textbf{60}, 309--324
  (1998).

\bibitem{mackowski_discrete_2002}
D.~W. Mackowski, \enquote{Discrete dipole moment method for calculation of the
  {T} matrix for nonspherical particles,} {\protect\JournalTitle{JOSA A}}
  \textbf{19}, 881--893 (2002).

\bibitem{loke_discrete-dipole_2011}
V.~L.~Y. Loke, M.~P. Mengüç, and T.~A. Nieminen, \enquote{Discrete-dipole
  approximation with surface interaction: {Computational} toolbox for
  {MATLAB},} {\protect\JournalTitle{Journal of Quantitative Spectroscopy and
  Radiative Transfer}} \textbf{112}, 1711--1725 (2011).

\end{thebibliography}


\begin{thebibliography}{1}
\newcommand{\enquote}[1]{``#1''}

\bibitem{zwillinger_table_2014}
D.~Zwillinger, ed., \emph{Table of {Integrals}, {Series}, and {Products}}
  (Academic Press, Amsterdam ; Boston, 2014), 8th ed.

\end{thebibliography}

%%%%%%%%%% If preparing manually:
% \begin{thebibliography}{1}
% \newcommand{\enquote}[1]{``#1''}

% \bibitem{Zhang:14}
% Y.~Zhang, S.~Qiao, L.~Sun, Q.~W. Shi, W.~Huang, L.~Li, and Z.~Yang,
%   \enquote{Photoinduced active terahertz metamaterials with nanostructured
%   vanadium dioxide film deposited by sol-gel method,}
%   {\protect\JournalTitle{Optics Express}} \textbf{22}, 11070--11078 (2014).

% \bibitem{OSA}
% {Optical Society}, \enquote{{OSA Publishing},}
%   \url{http://www.osapublishing.org}.

% \bibitem{FORSTER2007}
% P.~Forster, V.~Ramaswamy, P.~Artaxo, T.~Bernsten, R.~Betts, D.~Fahey,
%   J.~Haywood, J.~Lean, D.~Lowe, G.~Myhre, J.~Nganga, R.~Prinn, G.~Raga,
%   M.~Schulz, and R.~V. Dorland, \enquote{Changes in atmospheric consituents and
%   in radiative forcing,} in \enquote{Climate Change 2007: The Physical Science
%   Basis. Contribution of Working Group 1 to the Fourth assesment report of
%   Intergovernmental Panel on Climate Change,}  S.~Solomon, D.~Qin, M.~Manning,
%   Z.~Chen, M.~Marquis, K.~B. Averyt, M.~Tignor, and H.~L. Miler, eds.
%   (Cambridge University Press, 2007).

% \end{thebibliography}

\end{document}

\maketitle

\section{Analytical model for MSAD limit for small, axisymmetric particles}

Here, we calculate the limiting value as the delay time $\tau\rightarrow\infty$ for the mean-squared angular 
displacement (MSAD) of the symmetry axis of an axisymmetric particle (such as a dimer) strongly trapped 
in optical tweezers.
Let $\vec{u}_3$ be the symmetry axis of 
an axisymmetric particle at temperature $T$ that is fixed at the origin but undergoes rotational diffusion subject to an external
restoring torque given in spherical polar coordinates by
\be \label{eq:torque}
\vec{n} = -\kappa_r\sin(2\theta) \hat{\theta},
\ee
where $\kappa_r$ is a rotational stiffness.
The MSAD for $\vec{u}_3$ is defined as 
\be \label{eq:msad_defn}
\langle (\Delta \vec{u}_3)^2\rangle =  \langle \left| \vec{u}_3(t+\tau) - \vec{u}_3(t) \right|^2 \rangle.
\ee
As $\tau\rightarrow\infty$, $\vec{u}_3(t+\tau)$ and $\vec{u}_3(t)$ become statistically independent, and
the limiting value for $\langle (\Delta \vec{u}_3)^2 \rangle$
may be found using equilibrium statistical mechanics.
 
We focus on calculating the axis autocorrelation $\langle\vec{u}_3(t+\tau) \cdot \vec{u}_3(t) \rangle$ since
expansion of $\langle (\Delta \vec{u}_3)^2 \rangle$ (Eq.~\ref{eq:msad_defn}) leads to 
\be \label{eq:msad_expansion}
\langle (\Delta \vec{u}_3)^2\rangle = 2\left[1 - \langle \vec{u}_3(t+\tau) \cdot \vec{u}_3(t) \rangle \right].
\ee
Since $\vec{u}_3$ is a unit vector,
we hereafter refer to the axis autocorrelation as $\langle \cos\psi \rangle$, 
where $\psi$ is the angle subtended between $\vec{u}_3$ at $t$ and $\vec{u}_3$ at $t+\tau$.
Let the coordinates of $\vec{u}_3$ on the unit sphere be $(\theta, \phi)$ at time $t$ and
$(\theta', \phi')$ at time $t+\tau$.
Expressing the Cartesian components of $\vec{u}_3$ in spherical coordinates and computing the dot product leads to
\be \label{eq:cos_psi}
\cos\psi = \sin\theta\sin\theta'\left( \cos\phi\cos\phi' + \sin\phi\sin\phi'  \right) + \cos\theta\cos\theta'.
\ee

In thermal equilibrium, the system is governed by Boltzmann statistics.
Taking the system to have potential energy $U=0$ at $\theta=0$,
integration of the external torque (Eq.~\ref{eq:torque}) leads to  
\be \label{eq:potential_energy}
U = \kappa_r \sin^2\theta.
\ee
So, at time $t$, the probability of $\vec{u}_3$ lying 
within solid angle $\mathrm{d}\Omega$ near $(\theta, \phi)$ is given by
\be \label{eq:prob}
P(\theta, \phi) \mathrm{d}\Omega = \frac{\exp\left( -\beta \kappa_r \sin^2\theta \right)\,\mathrm{d}\Omega}{\iint \exp\left( -\beta \kappa_r \sin^2\theta \right) \, 
\mathrm{d}\Omega},
\ee
where $\beta \equiv 1/k_BT$, $\mathrm{d}\Omega = \sin\theta\,\mathrm{d}\phi\,\mathrm{d}\theta$,
and the integrals in the denominator run over the entire unit sphere.
If $\tau$ is sufficiently large that the orientation of $\vec{u}_3$ at time $t+\tau$ is independent of its orientation at time $t$, 
then the probability of $\vec{u}_3$ lying near $(\theta',\phi')$ at $t+\tau$ is similarly given by  
\be \label{eq:prob'}
P(\theta', \phi') \mathrm{d}\Omega' = \frac{\exp\left( -\beta \kappa_r \sin^2\theta '\right)\,\mathrm{d}\Omega'}{\iint \exp\left( -\beta \kappa_r \sin^2\theta' \right) \, 
\mathrm{d}\Omega'}.
\ee
(To calculate $\langle \cos \psi \rangle$ at an arbitrary value of $\tau$, we would instead need the transition probability of finding 
$\vec{u}_3$ at $(\theta', \phi')$ after a time interval $\tau$ given $\vec{u}_3$ initially at $(\theta,\phi)$.)
It follows that
\be \label{eq:large_integral}
\lim_{\tau\rightarrow\infty} \langle \cos\psi\rangle = \iint \cos \psi P(\theta, \phi) P(\theta', \phi')\,\mathrm{d}\Omega\,\mathrm{d}\Omega'.
\ee

Equation (\ref{eq:large_integral}) simplifies considerably because the probabilities (Eqs.~\ref{eq:prob}, \ref{eq:prob'}) 
have no explicit $\phi$ or $\phi'$ dependence. 
Since the first two terms of Eq.~(\ref{eq:cos_psi}) are proportional to $\cos\phi$ or 
$\sin\phi$, they vanish upon integration over $\phi$.
Thus, only the last term of Eq.~(\ref{eq:cos_psi}) contributes to $\langle \cos\psi\rangle$. 
Integrating over $\phi$ and $\phi'$,
\be \label{eq:big_integrals}
\lim_{\tau\rightarrow\infty}  \langle\cos\psi\rangle = 
\frac{\left(\int \cos\theta \exp\left( -\beta \kappa \sin^2\theta \right) \sin\theta\,\mathrm{d}\theta \right)
\left(\int \cos\theta' \exp\left( -\beta \kappa \sin^2\theta' \right) \sin\theta'\,\mathrm{d}\theta' \right) }{ \left( 
\int \exp\left( -\beta \kappa \sin^2\theta \right) \sin\theta\,\mathrm{d}\theta \right) \left(\int  \exp\left( -\beta \kappa \sin^2\theta' \right) \sin\theta'\,\mathrm{d}\theta' \right)} .
\ee

The limits of integration now require some attention. The integrals in the numerator of 
Eq.~(\ref{eq:big_integrals}) are proportional to $\langle \cos \theta \rangle$.
But the potential energy in Eq.~(\ref{eq:potential_energy}) implies that 
$\vec{u}_3$ is equally likely to be at $\theta$ or at  $\pi-\theta$.
Consequently, $\langle \theta \rangle = \pi/2$ and $\langle \cos\theta\rangle = 0$.
Instead, we note that in our simulations, the dimers are strongly trapped: $\vec{u}_3$ always lies near
$\theta = 0$ and never flips to lie near $\theta = \pi$.
Assuming that $\vec{u}_3$ is confined to the upper hemisphere, we can restrict 
the upper limits of integration to $\theta$ or $\theta' = \pi/2$.
Since the integrals over $\theta$ and $\theta'$ are independent, we then have
\be
\lim_{\tau\rightarrow\infty} \langle \cos \psi\rangle  = \left( \frac{I}{Z} \right)^2,
\ee
where
\be 
I \equiv \int_0^{\pi/2} \sin\theta\cos\theta \exp\left( -\beta \kappa_r \sin^2\theta \right) \,\mathrm{d}\theta
\ee
and
\be 
Z \equiv \int_0^{\pi/2} \sin\theta \exp\left( -\beta \kappa_r \sin^2\theta \right) \,\mathrm{d}\theta.
\ee

\begin{figure}[htbp]
\centering
\includegraphics{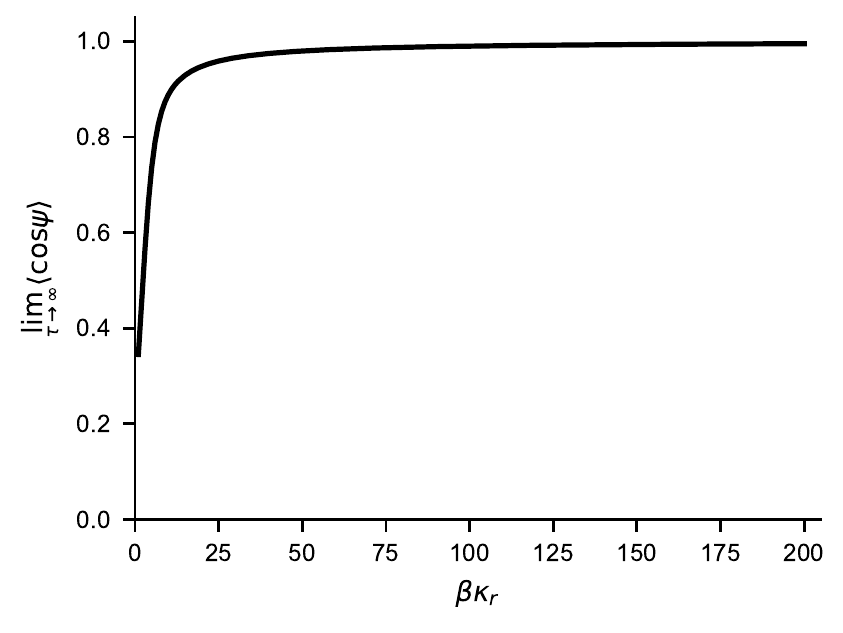}
\caption{\label{fig:betakappa}
Dependence of limiting value of $\langle \cos\psi\rangle$ [Eq.~(\ref{eq:main_result})]
on $\beta\kappa_r$. In the limit of increasing rotational stiffness, 
$\langle \cos \psi\rangle \rightarrow 1$.
}
\end{figure}

Evaluating $I$ by substitution leads to 
\be \label{eq:I1_result}
I = \frac{1}{2\beta\kappa_r}\left[ 1 - \exp(-\beta\kappa_r)  \right].
\ee
Evaluating $Z$ leads to \cite{zwillinger_table_2014}
\be \label{eq:Z_result}
Z = \frac{1}{2}\sqrt{\frac{\pi}{\beta\kappa_r}}\exp(-\beta\kappa_r) \mathrm{erfi}(\sqrt{\beta\kappa_r}),
\ee
where $\mathrm{erfi}(x)$ is the imaginary error function.
So, 
\be 
\lim_{\tau\rightarrow\infty} \langle \cos\psi \rangle = \frac{1}{\pi\beta\kappa_r}  \left( \frac{\exp(\beta\kappa_r) -1 }{\mathrm{erfi} 
(\sqrt{\beta\kappa_r})}  \right)^2.
\ee
For numerical calculations, it is convenient to express the result in terms of Dawson's integral
\be
F(x) \equiv \exp(-x^2) \int_0^x \exp(t^2)\,\mathrm{d}t,
\ee
where $\mathrm{erfi}(x) = \frac{2}{\sqrt{\pi}}\exp(x^2)F(x)$.
Thus, 
\be \label{eq:main_result}
\lim_{\tau\rightarrow\infty} \langle \cos\psi \rangle = \frac{1}{4\beta\kappa_r}  \left( \frac{\exp(\beta\kappa_r) -1 }{\exp(\beta\kappa_r) F(\sqrt{\beta\kappa_r}) }  \right)^2.
\ee
The dependence of Eq.~(\ref{eq:main_result}) on $\beta\kappa_r$ is shown in Fig.~\ref{fig:betakappa}.
Note that the limiting value approaches 1 as $\beta\kappa_r\rightarrow\infty$:
as the trap becomes increasingly stiff, $\vec{u}_3$ lies increasingly close to the $+z$ axis.

Finally, substitution of Eq.~(\ref{eq:main_result}) into Eq.~(\ref{eq:msad_expansion}) leads to
\be \label{eq:msad_limit}
\lim_{\tau\rightarrow\infty} \langle( \Delta \vec{u}_3)^2\rangle = 2\left[ 1 - \frac{1}{4\beta\kappa_r} 
\left( \frac{\exp(\beta\kappa_r) - 1}{\exp(\beta \kappa_r)F(\sqrt{\beta\kappa_r})}\right)^2    \right],
\ee
which is 
Eq.~(12) in the main text.
%Eq.~(\ref{M-eq:msad_limit}) in the main text.

\begin{figure}[htbp]
\centering
\includegraphics{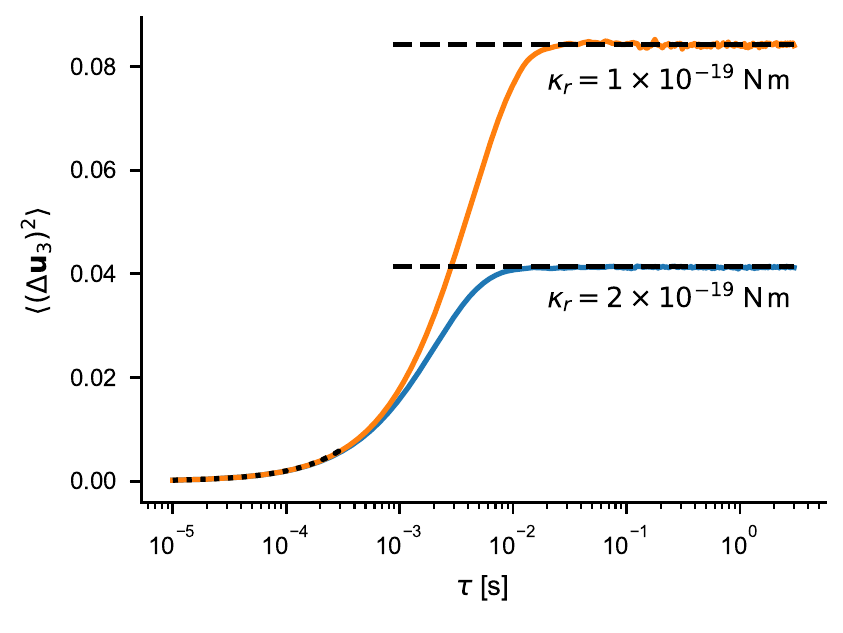}
\caption{\label{fig:sinusoid} $\vec{u}_3$ MSADs calculated from simulations of a particle 
at $T=\SI{295}{\kelvin}$ with 
$D_{r,\perp} = \SI{5}{\second^{-1}}$
and the labeled values of $\kappa_r$. Each MSAD is averaged from 5 trajectories. Dashed lines indicate predictions
of Eq.~(\ref{eq:msad_limit}), and the dotted line indicates the predictions of Eq.~(\ref{eq:diffusive_msad}).}
\end{figure}

To validate this result, we perform Brownian dynamics simulations in which the particle is fixed in place (with all elements of the
translational diffusion tensor identically 0) and experiences the external torque given by Eq.~\ref{eq:torque} 
(Fig.~\ref{fig:sinusoid}). The limiting values of the MSADs for $\vec{u}_3$ calculated from the simulated trajectories 
agree with the predictions of Eq.~(\ref{eq:msad_limit}).
The simulations also show that the short-time dynamics are governed by 
\be \label{eq:diffusive_msad}
\lim_{\tau\rightarrow 0} \langle (\Delta \vec{u}_3)^2 \rangle = 2\left[1 - \exp(-2D_{r,\perp}\tau)\right].
\ee
where $D_{r,\perp}$ is the element of the rotational diffusion tensor for rotations about $\vec{u}_1$ and
$\vec{u}_2$.

% Bibliography
\bibliography{bd}

%Manual citation list
%\begin{thebibliography}{1}
%\bibitem{Zhang:14}
%Y.~Zhang, S.~Qiao, L.~Sun, Q.~W. Shi, W.~Huang, %L.~Li, and Z.~Yang,
 % \enquote{Photoinduced active terahertz metamaterials with nanostructured
  %vanadium dioxide film deposited by sol-gel method,} Opt. Express \textbf{22},
  %11070--11078 (2014).
%\end{thebibliography}